\documentclass[a4paper,11pt]{article}
\usepackage{jheppub}
\usepackage[T1]{fontenc} 
\usepackage{slashed}
\usepackage[utf8x]{inputenc}
\usepackage{color}
\definecolor{red}{rgb}{1,0,0}
\usepackage[normalem]{ulem}
\usepackage{bm}
\def\siml{{\ \lower-1.2pt\vbox{\hbox{\rlap{$<$}\lower6pt\vbox{\hbox{$\sim$}}}}\ }}
\def\simg{{\ \lower-1.2pt\vbox{\hbox{\rlap{$>$}\lower6pt\vbox{\hbox{$\sim$}}}}\ }}

\allowdisplaybreaks

\title{CP asymmetry in heavy Majorana neutrino decays at finite temperature: the nearly degenerate case}

\author[a]{S. Biondini,}
\author[a,b]{N. Brambilla,}
\author[c]{M. A. Escobedo}
\author[a]{and A. Vairo}

\affiliation[a]{Physik-Department, Technische Universit\"{a}t M\"{u}nchen,\\ James-Franck-Str. 1, 85748 Garching, Germany}
\affiliation[b]{Institute for Advanced Study, Technische Universit\"{a}t M\"{u}nchen,\\ Lichtenbergstrasse 2 a, 85748 Garching, Germany}
\affiliation[c]{Institut de Physique Th\`eorique, Universit\'e Paris Saclay, CNRS, CEA,\\ F-91191 Gif-sur-Yvette, France}

\preprint{TUM-EFT 41/13}

\emailAdd{simone.biondini@tum.de}
\emailAdd{nora.brambilla@tum.de}
\emailAdd{miguel-angel.escobedo-espinosa@cea.fr}
\emailAdd{antonio.vairo@tum.de}

\abstract{In a model where Majorana neutrinos heavier than the electroweak scale 
couple to Standard Model Higgs bosons and leptons, we compute systematically thermal 
corrections to the direct and indirect CP asymmetries in the Majorana neutrino decays. 
These are key ingredients entering the equations that describe the thermodynamic evolution 
of the induced lepton-number asymmetry eventually leading to the baryon asymmetry in the universe.
We compute the thermal corrections in an effective field theory framework that assumes the temperature 
smaller than the masses of the Majorana neutrinos and larger than the electroweak scale,  
and we provide the leading corrections in an expansion of the temperature over the mass.
In this work, we consider the case of two Majorana neutrinos with nearly degenerate masses.}


\begin{document} 
\maketitle
\flushbottom

\section{Introduction}
Observations suggest that the number of baryons in the universe is 
different from the number of anti-baryons. The almost total absence of
antimatter on Earth, in our solar system and in cosmic rays indicates
that the universe is baryonically asymmetric. Indeed there are
observables to make this statement more quantitative. The baryon
asymmetry in the universe may be expressed in terms of the
baryon to photon ratio
\begin{equation}
\eta \equiv \frac{n_{B}-n_{\bar{B}}}{n_{\gamma}}=(6.21 \pm 0.16) \times 10^{-10} \, ,
\end{equation} 
where $n_{B}$, $n_{\bar{B}}$ and $n_{\gamma}$ are the number densities
of baryons, anti-baryons and photons respectively. Such a value comes
from accurate measurements of the anisotropies in the cosmic microwave
background~\cite{Larson:2010gs}. Consistent results come from
the comparison between the abundances of the light elements 
(D, $^{3}$He, $^{4}$He and $^7$Li), with the predictions of big bang nucleosynthesis~\cite{Iocco:2008va}. 
Such baryon asymmetry could be set as an initial condition for the universe evolution. 
However, it would require a high fine tuning and the initial baryon asymmetry
would be washed out during the inflationary period. This is why the scenario of a
dynamically generated baryon asymmetry is more appealing.

The dynamical generation of a baryon asymmetry in the context of quantum field theory
is called baryogenesis. One of the most attractive
and field theoretically consistent frameworks for baryogenesis is
via leptogenesis~\cite{Fukugita:1986hr}. In the original formulation, leptogenesis
requires a modest extension of the Standard Model (SM), namely, the
addition of right-handed neutrinos with large Majorana masses, far
above the electroweak scale $M_{W}$. The right-handed (sterile) neutrinos are
singlets under the SM gauge groups, whereas they are minimally coupled to the
SM particles via complex Yukawa couplings. These provide an additional
source of CP violation with respect to the one already present in the quark sector of the SM. 
In the standard picture, the heavy neutrinos are produced by thermal scatterings in the early universe and then decay
out of equilibrium either in SM leptons or anti-leptons in different
amounts due to the CP violating phases. Such an asymmetry in the
lepton sector is then partially reprocessed in a baryon asymmetry by sphaleron transitions in the SM~\cite{Kuzmin:1985mm}.

Majorana neutrino decays happen in a hot medium, namely the universe in its early stages.
Interactions with the medium modify the neutrino dynamics (thermal production rate, mass, ...) 
and affect the thermodynamic evolution of the lepton asymmetry.
The thermal production rate of right-handed neutrinos has been
studied in~\cite{Laine:2013lka} in the relativistic and ultra-relativistic regimes. 
The non-relativistic regime also turns out to be interesting for leptogenesis since it is conceivable that the CP
asymmetry is effectively generated when the temperature of the plasma
drops below the heavy-neutrino mass. In this regime the thermal
production rate for heavy Majorana neutrinos has been addressed in~\cite{Salvio:2011sf,Laine:2011pq}.

In~\cite{Biondini:2013xua} we used an effective field theory (EFT)
to describe the effective interactions between non-relativistic Majorana neutrinos and SM particles 
at a finite temperature $T$, assuming the following hierarchy of scales
\begin{equation}
M \gg T \gg M_{W} \, ,
\label{hiera}
\end{equation} 
where $M$ is the mass scale of the Majorana neutrinos.
In the temperature window \eqref{hiera} and in an expanding universe the heavy
neutrino is likely out of equilibrium, which is one of the Sakharov 
conditions necessary for generating a lepton asymmetry~\cite{Sakharov:1967dj}.  
In this paper, we study, in the same framework and under the same assumption, 
the thermal corrections to the CP asymmetry in the leptonic decays of heavy neutrinos,  
which is defined as
\begin{equation}
\epsilon_{I}=
\frac{\sum_{f} \Gamma(\nu_{R,I} \to \ell_{f} + X)-\Gamma(\nu_{R,I} \to \bar{\ell}_{f}+ X )  }
{\sum_{f} \Gamma(\nu_{R,I} \to \ell_{f} + X ) + \Gamma(\nu_{R,I} \to \bar{\ell}_{f}+ X)} \, .
\label{eq:adef}
\end{equation}
The sum runs over the SM lepton flavours, $\nu_{R,I}$ stands for the
$I$-th heavy right-handed neutrino species, $\ell_{f}$ is a SM lepton with flavour $f$ 
and $X$ stands for any other SM particle not carrying a lepton number. 
Another Sakharov condition necessary for baryogenesis is the occurrence of C and CP violating processes.
The quantity $\epsilon_{I}$ is a measure of the CP asymmetry generated by the decay of the $I$-th heavy neutrino, 
and we will refer to it in this way. Moreover, $\epsilon_I$ multiplied by the
corresponding neutrino number density enters the Boltzmann equations describing 
the thermodynamic evolution of the lepton-number asymmetry~\cite{Kolb:1979ui}. 
The quantity $\epsilon_{I}$ is also called unflavoured CP asymmetry because it does 
not distinguish between the different lepton flavour families. 
If the sum over the flavours is omitted in the numerator of \eqref{eq:adef}, 
then this defines what is called the flavoured CP asymmetry. 
We will discuss relevance of and compute the flavoured CP asymmetry in section~\ref{sec:flavour}.

\begin{figure}[ht]
\centering
\includegraphics[scale=0.585]{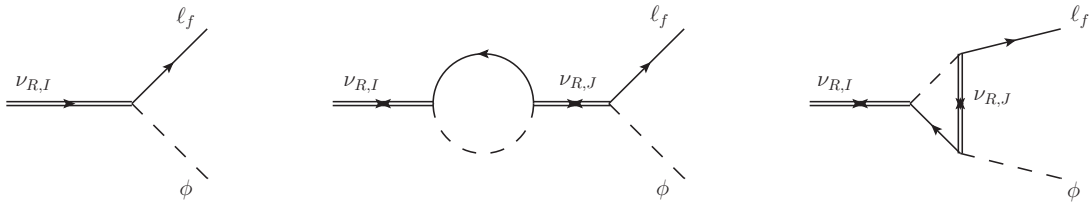}
\caption{From left to right: tree-level, and one-loop self-energy and vertex diagrams.
Double solid lines stand for heavy right-handed neutrino propagators,
solid lines for lepton propagators and dashed lines for Higgs boson propagators.
The neutrino propagator with forward arrow corresponds to $\langle 0| T(\psi \bar{\psi}) |0\rangle$,  
whereas the neutrino propagators with forward-backward arrows correspond to $\langle 0| T(\psi \psi) |0\rangle$ 
or $\langle 0| T(\bar{\psi} \bar{\psi})|0 \rangle$, see appendix \ref{AppendixA}.}
\label{drind} 
\end{figure}

The CP asymmetry is originated from the interference between the tree-level and the one-loop self-energy and vertex diagrams shown in figure~\ref{drind}. 
The contribution from the interference with the self-energy diagram is often called indirect 
contribution, while the one from the interference with the vertex diagram is called direct contribution. 
The relative importance of the indirect and direct contributions for the CP asymmetry depends on the heavy-neutrino mass spectrum. 
For example, the vertex contribution is half of the self-energy contribution in the hierarchical case,
when the mass of one species of neutrinos is much lighter than the others~\cite{Liu:1993tg,Covi:1996wh}. 
The situation is different when two heavy neutrinos are nearly degenerate in mass. 
In this case, the self-energy diagram can develop a resonant enhancement that is related to a mixing
phenomenon similar to the one found in kaon physics, as originally proposed in~\cite{Flanz:1996fb}. 
An analysis from first principles has been carried out in~\cite{Buchmuller:1997yu,Garny:2011hg,Garbrecht:2011aw}. 
The main phenomenological outcome is that the scale of the heavy right-handed neutrino masses 
can be lowered down to energy scales of $\mathcal{O}$(TeV)~\cite{Pilaftsis:2003gt}. 
However, also the nearly degenerate case may comprise situations in which both the
vertex and self-energy diagrams contribute to the CP asymmetry with a similar magnitude~\cite{Pilaftsis:1998pd}, 
namely when the peculiar condition for resonant leptogenesis is not met.

A thermal treatment of the lepton-number asymmetry in the resonant case, 
i.e., when the mass difference of the heavy neutrinos is of the order of magnitude of their decay widths, 
can be found for instance in~\cite{Garny:2011hg}, where the Boltzmann equations 
are superseded by the quantum version known as Kadanoff--Baym equations.  
The lepton-number asymmetry has been also considered for a generic heavy-neutrino mass spectrum, 
e.g., in~\cite{Covi:1997dr,Giudice:2003jh,Garny:2010nj,Anisimov:2010dk,Kiessig:2011fw} within different approaches. 
The thermal effects considered include using thermal masses for the Higgs boson and leptons and 
taking into account thermal distributions for the Higgs boson and leptons as decay products of the heavy Majorana neutrinos.

In this work, we aim at treating systematically thermal effects to the CP asymmetry \eqref{eq:adef} in the non-relativistic regime specified by \eqref{hiera}. 
These effects lead to corrections in terms of series in the SM couplings and in $T/M$ 
in the same way as they do for the heavy Majorana neutrino production rate~\cite{Salvio:2011sf,Laine:2011pq}.
We will derive such thermal corrections for the case of two Majorana neutrinos with nearly degenerate masses, 
i.e., we will assume a mass splitting much smaller than $M$.
We will not specify, however, the relation between the mass splitting and the widths. 
Hence our treatment includes, but is not limited to, the case when the mass splitting is of the order of the widths. 
The CP asymmetry is proportional to the imaginary parts of the Majorana neutrino Yukawa couplings.
We note that in the exact degenerate case the CP phases can be rotated away leading to purely
real Yukawa couplings, and, therefore, to a vanishing CP asymmetry~\cite{Buchmuller:1997yu}.
We will discuss the hierarchical case elsewhere~\cite{Biondinihier}. 

Systems with two nearly degenerate heavy Majorana neutrinos are characterized by one large scale: $M$.
They may be treated in the non-relativistic EFT framework introduced in~\cite{Biondini:2013xua}. 
There are some advantages in such an approach.
First, the power counting of the EFT allows to assess a priori the size of the 
different corrections to the CP asymmetry optimizing the calculation. 
Moreover, the calculation, which would involve three-loop diagrams in a relativistic thermal field theory, 
can be split into a simpler two-step evaluation. Similarly to what is done in~\cite{Biondini:2013xua} for the thermal production rate, 
the first step consists, by power counting, in the evaluation of the imaginary parts of the Wilson coefficients
of some dimension five operators in the EFT. 
The Wilson coefficients encode the physics from the mass scale, $M$.
Since $M\gg T$, they may be computed setting the temperature to zero.
In our case, this step consists in computing electroweak two-loop cut diagrams in vacuum.  
The second step requires the computation of a simple thermal one-loop diagram in the EFT.
The disadvantage of the approach consists in being limited to temperatures for which \eqref{hiera} holds.

The paper is organized as follows. In section~\ref{sec:rev} and appendix~\ref{AppendixA} we review
the basic set-up of the EFT for non-relativistic Majorana neutrinos. 
In section~\ref{sec:zeroT} we re-derive the zero temperature direct CP asymmetry from the vertex diagram and relate it to the EFT. 
In section~\ref{sec:finiteT} we match the relevant dimension-five operators of the EFT at two loops. 
The detailed calculation can be found in appendix~\ref{AppendixB}.
The leading thermal corrections to the direct CP asymmetry are computed in section~\ref{sec:direct}
and  the leading thermal corrections to the indirect CP asymmetry in section~\ref{sec:indirect}.
In section~\ref{sec:flavour}, we extend our study to the flavoured CP asymmetry, some of whose contributions are evaluated at the end of appendix~\ref{AppendixB}. 
We discuss general issues related to the convergence of the relativistic expansion in appendix~\ref{AppendixD}.
Finally, conclusions are drawn in section~\ref{sec:concl}.

\section{EFT for non-relativistic Majorana neutrinos}
\label{sec:rev}
We start by specifying our model of new physics. 
We work within a conservative extension of the SM that consists in adding right-handed neutrinos to the SM particle content. 
To generate a non-vanishing CP asymmetry \eqref{eq:adef} at least two different neutrino species have to be added. 
In the following, we will consider only two heavy neutrinos and assume that they have masses above the electroweak scale.  
In the case that right-handed neutrinos are represented by Majorana fermion fields, 
the Lagrangian may be written as follows~\cite{Fukugita:1986hr} (we adopt some of the notation of~\cite{Asaka:2006rw}):
\begin{equation}
\mathcal{L}=\mathcal{L}_{\rm{SM}} + \frac{1}{2} \bar{\psi}_{I} i \slashed{\partial}  \psi_{I}  
- \frac{M_{I}}{2} \bar{\psi}_{I}\psi_{I} - F_{f I}\bar{L}_{f} \tilde{\phi} P_{R}\psi_{I}  - F^{*}_{f I}\bar{\psi}_{I} P_{L} \tilde{\phi}^{\dagger}  L_{f} \, ,
\label{eq3}
\end{equation} 
where $\psi_{I}=\nu_{R,I}+\nu^{c}_{R,I}$ is the Majorana field
comprising the right-handed neutrino $\nu_{R,I}$ of type $I$ ($I=1,2$)
and mass $M_{I}$; $\mathcal{L}_{\rm{SM}}$ is the SM Lagrangian with unbroken SU(2)$_L\times$U(1)$_Y$ gauge symmetry (see \eqref{SMlag}), 
$\tilde{\phi}=i \sigma^{2} \, \phi^*$ embeds the SM Higgs doublet, 
$L_{f}$ is the SM lepton doublet of flavour $f$, $F_{fI}$ is a complex Yukawa coupling, 
and the right-handed and left-handed projectors are denoted by $P_R = (1 + \gamma^5)/2$ 
and $P_L = (1 - \gamma^5)/2$ respectively. 
We consider the nearly degenerate case where $M_2-M_1 \ll M_1 \sim M_2$.
We call neutrino of type 2 the heaviest of the two neutrinos, and,  
for further use, we define $0 < \Delta \equiv M_2-M_1$ and $M\equiv M_1$.

We will compute the thermal modification induced to the CP asymmetry of the Majorana neutrino decays 
by a plasma of SM particles at a temperature $T$ under the conditions $M\gg T \gg M_{W}$ and $M \gg \Delta$. 
We exploit the hierarchy $M \gg T$ by performing the calculation in two steps.
First we integrate out momentum and energy modes of order $M$ from the 
fundamental Lagrangian~\eqref{eq3} and replace it by a suitable effective field theory aimed 
at describing the non-relativistic dynamics of the Majorana neutrinos.
The EFT is organized as an expansion in operators of increasing dimension suppressed by powers of $1/M$.
The Wilson coefficients of the operators encode the high-energy modes of the fundamental theory and can be 
evaluated by setting $T=0$.  Then we compute thermal corrections to the Majorana neutrino leptonic widths 
as thermal averages weighted by the partition function of the EFT.
The EFT for non-relativistic Majorana neutrinos was introduced and discussed in the case 
of one right-handed neutrino generation in~\cite{Biondini:2013xua}.
The framework here is very similar, the only difference being that we deal with two generations of neutrinos instead of one. 
The EFT Lagrangian up to operators of dimension five is
\begin{equation}
\mathcal{L}_{\rm{EFT}}=\mathcal{L}_{\rm{SM}} 
+ \bar{N}_I \left( i v \cdot \partial -\delta M_I \right) N_I +
\frac{i\Gamma^{T=0}_{IJ}}{2}\bar{N}_I N_J+\frac{a_{IJ}}{M_I}\bar{N}_IN_J \phi^{\dagger}\phi + \dots,
\label{eq:efflag}
\end{equation}
where $N_{I}$ is the field describing the low-energy modes of the $I$-th non-relativistic Majorana neutrino, 
$\delta M_1=0$, $\delta M_2=\Delta$, $\Gamma^{T=0}_{IJ}$ is the decay matrix at $T=0$ 
and $a_{IJ}$ are the Wilson coefficients of the dimension-five operators $\bar{N}_IN_J \phi^{\dagger}\phi$
describing the interaction of the Majorana neutrinos with the Higgs doublet of the SM. 
These are the only operators of dimension five that give thermal corrections to the neutrino widths and masses. 
The dots in \eqref{eq:efflag} stand for higher-order operators that contribute with subleading 
corrections and that are beyond the accuracy of this work. In particular, thermal corrections induced by gauge bosons, 
leptons and (heavy) quarks turn out to be subleading.\footnote{
Subleading refers here to corrections that are parametrically suppressed by $(T/M)^2$ with respect to those calculated.
Large differences in the size of the SM couplings may however alter the numerical relevance of the different corrections.
Further considerations can be found in the conclusions.}
The natural dynamical scale of the EFT Lagrangian is the temperature, $T$. Since $T$ is larger than the electroweak 
scale, $\mathcal{L}_{\rm{SM}}$ is still the SM Lagrangian with unbroken SU(2)$_L\times$U(1)$_Y$ gauge symmetry. 

The Lagrangian \eqref{eq:efflag} has been obtained by integrating out the mass $M=M_1$ from the Lagrangian~\eqref{eq3}; 
$\delta M_2=\Delta \ll M$ is the residual mass of the neutrino of type 2. 
In~\eqref{eq:efflag} and in the rest of the paper, masses are understood as on-shell masses,
as it is typical of non-relativistic EFTs, which implies that off-diagonal elements 
of the mass matrix vanish; moreover, in the diagonal terms we will neglect terms that would contribute 
to the CP asymmetry at order $F^6$ or smaller~\cite{Kniehl:1996bd,Anisimov:2005hr}. 
Off-diagonal elements do not vanish for the absorbtive parts $i\Gamma^{T=0}_{IJ}/2$. The specification $T=0$ recalls that they are computed at $T=0$.
Finally, the Lagrangian \eqref{eq:efflag} has been written in a reference frame where the Majorana neutrinos 
have momentum $M v^\mu$ ($v^2=1$) up to a residual momentum that is much smaller than $M$.
In the following, we will assume that the thermal bath of SM particles is comoving with the 
Majorana neutrinos. A convenient choice of the reference frame is the rest frame $v^{\mu}=(1,\vec{0})$.

In the introduction, we have distinguished between indirect and direct CP asymmetry, 
the distinction being based on the leading-order processes shown in figure~\ref{drind}.
In this paper, we extend that distinction beyond leading order by calling contributions to the indirect 
CP asymmetry, $\Delta\Gamma_{I,{\rm indirect}}$, those that show the phenomenon of resonant enhancement, i.e., 
a large enhancement of the asymmetry when $\Delta$ is of the order of the largest 
between the neutrino width difference and the mixing vertices.
In the framework of a strict perturbative expansion in the Yukawa couplings, such a behaviour is induced 
by Feynman diagrams (like the second of figure~\ref{drind}) becoming singular in the limit $\Delta\to 0$, 
which signals a break down of the expansion in that limit. 
The singularity is eventually removed by resumming certain classes of diagrams, 
like those responsible for the width and/or the mixing of the different neutrinos.
Viceversa, we call contributions to the direct CP asymmetry, $\Delta\Gamma_{I,{\rm direct}}$, those that 
do not exhibit this phenomenon. Order by order in an expansion in the Yukawa couplings, 
Feynman diagrams that contribute to the direct CP asymmetry are not singular in the limit $\Delta\to 0$.
The CP asymmetry is the sum of these two kind of contributions: 
\begin{eqnarray}
\sum_{f} \Gamma(\nu_{R,I} \to \ell_{f} + X)-\Gamma(\nu_{R,I} \to \bar{\ell}_{f}+ X ) &=& \Delta\Gamma_{I,{\rm direct}} + \Delta\Gamma_{I,{\rm indirect}} \,.
\label{asysec2}
\end{eqnarray}

The term $\Delta\Gamma_{I,{\rm direct}}$ includes all contributions to the CP asymmetry that originate from single operators in the EFT 
and all contributions that come from mixing of operators in the EFT that do not show the phenomenon of resonant enhancement. 
Concerning the first class of contributions, at the accuracy of the Lagrangian~\eqref{eq:efflag} 
there are only dimension 3 and 5 operators that may have imaginary Wilson coefficients.
Concerning the second class of contributions, we will denote them $\Delta\Gamma_{I,{\rm direct}}^{\rm mixing}$.
At the order we are working, the only relevant contribution of this kind affects the heavier Majorana neutrino of type 2 
and will be computed in section~\ref{sec:direct2}.
Hence, $\Delta\Gamma_{I,{\rm direct}}$ reads 
\begin{equation}
\Delta\Gamma_{I,{\rm direct}} = \left(\Gamma^{\ell,T=0}_{II}-\Gamma^{\bar{\ell},T=0}_{II}\right) 
+ \left(\Gamma^{\ell,T}_{II,{\rm direct}}-\Gamma^{\bar{\ell},T}_{II,{\rm direct}}\right) 
+ \Delta\Gamma_{I,{\rm direct}}^{\rm mixing}\,,
\label{asysec2bis}
\end{equation}
with
\begin{equation}
\Gamma^{\ell,T}_{II,{\rm direct}} = \frac{2}{M}{\rm Im}\,a_{II}^\ell \,\langle \phi^\dagger(0)\phi(0)\rangle_T,
\qquad 
\Gamma^{\bar{\ell},T}_{II,{\rm direct}} = \frac{2}{M}{\rm Im}\,a_{II}^{\bar\ell} \,\langle \phi^\dagger(0)\phi(0)\rangle_T,
\label{asysec3}
\end{equation}
where the subscripts $\ell$ and $\bar{\ell}$ isolate the leptonic and anti-leptonic contributions.
The first term in the right-hand side of~\eqref{asysec2bis}, $\Gamma^{\ell,T=0}_{II}-\Gamma^{\bar{\ell},T=0}_{II}$,  
is the zero temperature contribution to the direct CP asymmetry, which we will compute in section~\ref{sec:zeroT}.
The second term, $\Gamma^{\ell,T}_{II,{\rm direct}}-\Gamma^{\bar{\ell},T}_{II,{\rm direct}}$,
isolates the dominant thermal correction to the direct CP asymmetry, which will be the main subject of the paper. 

In equation \eqref{asysec3} the thermal dependence is encoded in the Higgs thermal condensate $\langle \phi^\dagger(0)\phi(0)\rangle_T$, 
which at leading order reads
\begin{equation}
\langle \phi^\dagger(0)\phi(0)\rangle_T  = \frac{T^2}{6}.
\label{higgscondensate}
\end{equation}
The relative size of the thermal correction to the direct CP asymmetry is therefore $T^2/M^2$.
High-energy contributions induced by loops with momenta of the order of the neutrino mass 
are encoded in the Wilson coefficients $a_{II}^\ell$ and $a_{II}^{\bar{\ell}}$. 
Since the condensate is real, to compute the widths we need the imaginary parts of $a_{II}^\ell$ and $a_{II}^{\bar{\ell}}$. 
Their expressions, at order $F^2$ in the Yukawa couplings, can be easily inferred from~\cite{Biondini:2013xua}
(see also appendix~\ref{appcutting}) and the result reads
\begin{equation}
{\rm Im}\, a_{II}^\ell = {\rm Im}\, a_{II}^{\bar\ell} = -\frac{3}{16\pi}|F_{I}|^2\lambda.
\label{agen}
\end{equation}
The coupling $\lambda$ is the four-Higgs coupling. 
We have defined $|F_{I}|^2 \equiv \sum_{f} F_{f I} F^{*}_{f I}$ and, for further use, $F_{J}F_{I}^{*} \equiv \sum_f F_{fJ}F_{fI}^{*}$.

A necessary condition to produce a CP asymmetry, i.e., to get a non-vanishing difference from a final state 
with a lepton and one with an anti-lepton, is for ${\rm Im}\,a_{II}^\ell$ and ${\rm Im}\,a_{II}^{\bar{\ell}}$
to be sensitive to the phases of the Yukawa couplings $F_{fI}$.
At order $F^2$, ${\rm Im}\, a_{II}^\ell$ and ${\rm Im}\, a_{II}^{\bar\ell}$ are not.
Hence, to produce a non-vanishing direct CP asymmetry, one needs to compute at least corrections of order $F^4$. 
In fact, corrections proportional to $(F_{1}F^{*}_{2})^2$ are sensitive to the phases of the Yukawa couplings.
From the optical theorem the imaginary part of a two-loop diagram proportional to $(F_{1}F^{*}_{2})^2$ may be 
understood as the interference between a tree-level and a one-loop amplitude developing an imaginary part.

In section \ref{sec:finiteT} and appendix \ref{AppendixB}, we will evaluate the diagrams contributing to ${\rm Im}\,a_{II}^\ell$ and ${\rm Im}\,a_{II}^{\bar{\ell}}$ 
at order $F^4$ in the Yukawa couplings and up to first order in the SM couplings. 
This will be done by computing in the fundamental theory~\eqref{eq3}, at $T=0$,
two-loop amplitudes with two external Majorana neutrinos and two external Higgs particles and by matching 
them to the corresponding $a_{11}$ and $a_{22}$ vertices in the EFT. Out of all diagrams, we will 
select only those sensible to a CP phase, i.e., those involving the interference of Majorana neutrinos of 
type 1 with Majorana neutrinos of type 2. We will compute the imaginary parts of those diagrams.
It will be convenient to use cutting rules, since cuts through lepton propagators select neutrino 
decays into leptons, whereas cuts through anti-lepton propagators select decays into anti-leptons.
We restrict to cuts that separate the diagrams into a tree-level part and a one-loop part. 
As we will see in the next section, in order to contribute to the  CP asymmetry 
the remaining one-loop part has to produce a complex phase. Therefore the only diagrams that contribute are 
the ones whose one-loop part can, in turn, be cut into two tree-level diagrams.

The term $\Delta\Gamma_{I,{\rm indirect}}$ in \eqref{asysec2} contains all contributions that exhibit resonant enhancement.
We can further distinguish them in zero temperature contributions, 
$\Gamma^{\ell,T=0}_{II,{\rm indirect}}-\Gamma^{\bar{\ell},T=0}_{II,{\rm indirect}}$, and thermal contributions, 
$\Gamma^{\ell,T}_{II,{\rm indirect}}-\Gamma^{\bar{\ell},T}_{II,{\rm indirect}}$. They will be computed in  section~\ref{sec:indirect}. 
Clearly, an indirect CP asymmetry can only originate from the mixing of operators in the EFT.
While $\Gamma^{\ell,T=0}_{II}-\Gamma^{\bar{\ell},T=0}_{II}$ and $\Gamma^{\ell,T}_{II,{\rm direct}}-\Gamma^{\bar{\ell},T}_{II,{\rm direct}}$
depend only on the diagonal elements $\Gamma^{T=0}_{II}$ and $a_{II}$, 
contributions from the mixing will depend crucially on the off-diagonal elements of $\Gamma^{T=0}_{IJ}$ and $a_{IJ}$ too.

\section{Matching $\Gamma^{T=0}_{II}$: direct asymmetry at zero temperature}
\label{sec:zeroT}
The direct CP asymmetry \eqref{asysec2bis} depends on the Wilson coefficients $\Gamma^{T=0}_{II}$ and $a_{II}$ of~\eqref{eq:efflag}. 
In this section we compute the leptonic, $\Gamma^{\ell,T=0}_{II}$, and anti-leptonic, $\Gamma^{\bar{\ell},T=0}_{II}$, components of $\Gamma^{T=0}_{II}$.
In so doing we re-derive the expression for the direct CP asymmetry at zero temperature~\cite{Fukugita:1986hr}. 
Considerations made here will be used in the next section to select the parts of the Wilson coefficients 
${\rm Im}\,a_{II}^\ell$ and ${\rm Im}\,a_{II}^{\bar{\ell}}$ relevant for the thermal corrections to the direct CP asymmetry.

\begin{figure}[ht]
\centering
\includegraphics[scale=0.55]{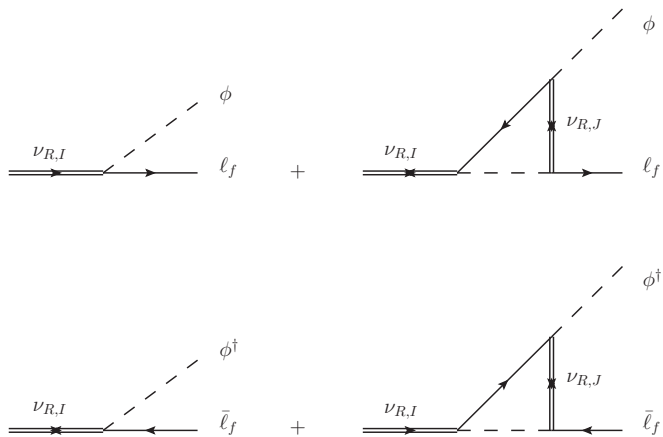}
\caption{Tree-level and one-loop diagrams contributing to the direct CP asymmetry.
The subscript $I$ stands either for 1 or 2. 
The first and second raws show decays into leptons and anti-leptons respectively.}
\label{Fig1} 
\end{figure}

\begin{figure}[ht]
\centering
\includegraphics[scale=0.45]{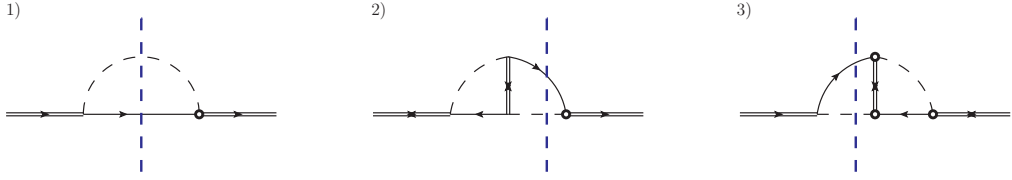}
\caption{One-loop and two-loops self-energy diagrams in the fundamental theory \eqref{eq3} 
contributing to the decay of a heavy Majorana neutrino into leptons.
Vertical blue dashed lines are the cuts selecting a final state made of a Higgs boson and a lepton.
Circled vertices and propagators are defined in appendix~\ref{appcutting}.}
\label{Fig2} 
\end{figure}

We start considering the decay of a heavy right-handed neutrino of type 1, $\nu_{R,1}$, into leptons. 
Up to one loop the amplitude has the following form (see the two upper diagrams in figure~\ref{Fig1} that display only direct 
contributions): 
\begin{equation}
\mathcal{M}(\nu_{R,1} \rightarrow \ell_{f}+X) =  A \left[ F_{f  1} + \sum_{J} (F^{*}_{ f' 1} F_{f'  J})F_{f  J} \, B \right] ,
\label{b2}
\end{equation} 
where $A$ and $B$ are functions that parameterize the amplitude at tree-level and one-loop respectively. 
We obtain the total decay width into leptons by squaring the amplitude and summing over the lepton flavours. 
Up to $\mathcal{O}(F^4)$ it reads
\begin{eqnarray}
&& \sum_{f} \Gamma(\nu_{R,1} \rightarrow \ell_f + X) =  |A|^2\left[ |F_{1}|^2 
+ \sum_{J} \left( (F^{*}_{1}F_{J})^2 \, B + ( F_{1} F^{*}_{J})^2 \, B^{*} \right) \right]  
\nonumber \\
&& \hspace{1.2cm}
=  |A|^2 \left\lbrace  |F_{1}|^2    + \sum_{J} \left(   2 \, {\rm{Re}}(B){\rm{Re}} \left[ (F_{1}^{*}F_{J})^2\right] 
-2 \, {\rm{Im}}(B) {\rm{Im}} \left[ (F_{1}^{*}F_{J})^2\right] \right)  \right\rbrace .
\label{b3}
\end{eqnarray} 
We may write similar relations for the decay into anti-leptons:
\begin{equation}
\mathcal{M}(\nu_{R,1} \rightarrow \bar{\ell_{f}} + X) =  A \left[ F^{*}_{f 1} + \sum_{J} (F_{f'1} F^{*}_{f' J})F^{*}_{fJ} \, C \right] ,
\label{b5}
\end{equation}
and 
\begin{eqnarray}
&& \sum_{f} \Gamma(\nu_{R,1} \rightarrow \bar{\ell}_f+X) = |A|^2\left[ |F_{1}|^2 
+ \sum_{J} \left( (F^{*}_{1}F_{J})^2 \, C^{*} + (F_{1}F^{*}_{J})^2 \, C \right) \right]  
\nonumber \\
&& \hspace{1.2cm}
=  |A|^2 \left\lbrace  |F_{1}|^2    + \sum_{J} \left( 2 \, {\rm{Re}}(C){\rm{Re}} \left[ (F_{1}^{*}F_{J})^2\right] +2 
\, {\rm{Im}}(C) {\rm{Im}} \left[ (F_{1}^{*}F_{J})^2\right] \right) \right\rbrace ,
\label{b6}
\end{eqnarray} 
where $C$ is the analogous of $B$ in~(\ref{b2}). 
The CP asymmetry~(\ref{eq:adef}), due to the decay of $\nu_{R,1}$, is then 
\begin{equation}
\epsilon_{1}= \sum_{J}  \, \frac{\left( {\rm{Re}}(B)-{\rm{Re}}(C) \right) {\rm{Re}} \left[ (F_{1}^{*}F_{J})^2\right] 
- \left( {\rm{Im}}(B)+{\rm{Im}}(C) \right) {\rm{Im}} \left[ (F_{1}^{*}F_{J})^2 \right] }{|F_{1}|^2} \, .
\label{cpzero}
\end{equation}
The functions $A$, $B$ and $C$ can be computed by cutting one and two-loop diagrams contributing to the propagator of a neutrino of type~1:
\begin{equation}
-i \left. \int d^{4}x \, e^{ip\cdot x} \, 
\langle \Omega | T \left( \psi_{1}^{\mu}(x) \bar{\psi}_{1}^{\nu}(0) \right) | \Omega \rangle \right|_{p^\alpha =(M + i\epsilon,\vec{0}\,)} \, ,
\label{matrixFund}
\end{equation}
where $|\Omega\rangle$ is the ground state of the fundamental theory 
and where we have chosen the rest frame $v^{\alpha}=(1,\vec{0})$, so that the incoming momentum is $p^\alpha =(M,\vec{0}\,)$.
Diagrams with cuts through lepton propagators contribute to $A$ and $B$ (see figure~\ref{Fig2}), while diagrams with 
cuts through anti-lepton propagators contribute to $A$ and $C$.
An analogous equation to \eqref{cpzero} holds for $\epsilon_{2}$.

We consider the in-vacuum diagrams in figure~\ref{Fig2} for incoming and outgoing neutrinos of type~1. 
The cuts select the contribution to the width into leptons (for details see appendix~\ref{appcutting}).
We call $\mathcal{D}^{\ell}_{1}$, $\mathcal{D}^{\ell}_{2}$ and $\mathcal{D}^{\ell}_{3}$ respectively
the diagrams shown in figure~\ref{Fig2} with amputated external legs.
The quantity ${\rm Im}\left[-i(\mathcal{D}^{\ell}_{1}+\mathcal{D}^{\ell}_{2}+\mathcal{D}^{\ell}_{3})\right]$ provides 
$\delta^{\mu\nu}\,\sum_f \Gamma(\nu_{R,1} \rightarrow \ell_f+X)/2$ at $T=0$ in the fundamental theory \eqref{eq3}, 
which matches $\delta^{\mu\nu}\,\Gamma^{\ell,T=0}_{11}/2$ in the EFT~\eqref{eq:efflag}.
The quantities $\Gamma^{\ell,T=0}_{II}$ and $\Gamma^{\bar{\ell},T=0}_{II}$ are the leptonic and anti-leptonic components of $\Gamma^{T=0}_{II}$ respectively. 
At leading order $\Gamma_{II}^{T=0}=\Gamma_{II}^{\ell,T=0}+\Gamma_{II}^{\bar{\ell},T=0}$. 
An explicit calculation up to order $\Delta/M$ gives
\begin{eqnarray}
&&
\delta^{\mu\nu}\,\frac{\Gamma^{\ell,T=0}_{11}}{2} = 
{\rm{Im}}\left[ -i (\mathcal{D}^{\ell}_{1} +\mathcal{D}^{\ell}_{2} +\mathcal{D}^{\ell}_{3})  \right]  = 
\nonumber \\
&& \hspace{1cm}
 \delta^{\mu \nu} \frac{M}{16 \pi}\left\lbrace  \frac{|F_1|^2}{2}  - \frac{\sum_{J=1}^2 {\rm{Re}}\left[ (F_1^* F_J)^2\right]}{(4 \pi)^2}  
\left[ \left( 1-\frac{\pi^2}{6}\right) + \left( 1-\frac{\pi^2}{12} -4 \ln 2\right) \frac{\Delta}{M} \right]  \right.  
\nonumber \\
&& \hspace{1cm}
\left. -\frac{\sum_{J=1}^2 {\rm{Im}}\left[ (F_1^* F_J)^2\right] }{16 \pi} \left[   (-1 +2 \ln 2) + (-3 + 4 \ln 2)\frac{\Delta}{M}   \right]  \right\rbrace .
\label{eq15}
\end{eqnarray}
The sum over $J$ comes from the flavour of the intermediate Majorana neutrino exchanged in the two-loop diagrams, 
clearly $\sum_J {\rm{Im}} (F_1^* F_J)^2 =  {\rm{Im}} (F_1^* F_2)^2$.
We have not considered cuts through the intermediate neutrino, which would correspond to neutrino transitions 
involving the emission of a lepton and an anti-lepton, because they do not contribute to the CP asymmetry.

The analogous calculation for $\sum_f \Gamma(\nu_{R,1} \rightarrow \bar{\ell}_f+X)$ at $T=0$ in the fundamental theory, 
which matches $\Gamma^{\bar{\ell},T=0}_{11}$ in the EFT, requires the calculation of the one-loop diagram with a virtual anti-lepton and 
the two-loop diagrams shown in figure~\ref{Fig2} but with cuts through anti-lepton propagators. Up to order $\Delta/M$, we obtain 
\begin{eqnarray}
&&
\delta^{\mu\nu}\,\frac{\Gamma^{\bar{\ell},T=0}_{11}}{2} = 
{\rm{Im}}\left[ -i (\mathcal{D}^{{\bar\ell}}_{1} +\mathcal{D}^{{\bar\ell}}_{2} +\mathcal{D}^{{\bar\ell}}_{3})  \right]  = 
\nonumber \\
&& \hspace{1cm}
\delta^{\mu \nu} \frac{M}{16 \pi}\left\lbrace  \frac{|F_1|^2}{2}  - \frac{\sum_{J=1}^2 {\rm{Re}}\left[ (F_1^* F_J)^2\right] }{(4 \pi)^2}  
\left[ \left( 1-\frac{\pi^2}{6}\right) + \left( 1-\frac{\pi^2}{12} -4 \ln 2\right) \frac{\Delta}{M} \right]  \right.  
\nonumber \\
&& \hspace{1cm}
\left. +\frac{\sum_{J=1}^2 {\rm{Im}}\left[ (F_1^* F_J)^2 \right] }{16 \pi} \left[   (-1 +2 \ln 2) + (-3 + 4 \ln 2)\frac{\Delta}{M}   \right] \right\rbrace .
\label{eq16}
\end{eqnarray}
The right-hand side of \eqref{eq16} differs from the right-hand side of \eqref{eq15} only for the sign of the term proportional to ${\rm{Im}}\left[ (F_1^* F_J)^2 \right]$. 
It is precisely this term that originates the CP asymmetry.

From \eqref{eq15} and \eqref{eq16} it follows:
\begin{eqnarray}
\Gamma^{\ell,T=0}_{11}-\Gamma^{\bar{\ell},T=0}_{11} &=& 
- \frac{M}{64 \pi^2} \left[   (-1 +2 \ln 2) + (-3 + 4 \ln 2)\frac{\Delta}{M}   \right] {\rm{Im}}\left[ (F_1^* F_2)^2 \right] ,  
\label{EQ17}\\
\Gamma^{T=0}_{11} &=& \Gamma^{\ell,T=0}_{11} + \Gamma^{\bar{\ell},T=0}_{11} = \frac{M}{8\pi}|F_1|^2,
\label{Gamma1T0}
\end{eqnarray}
where in the last line we have neglected terms of order $F^4$.
The direct CP asymmetry at $T=0$ for the leptonic decay of a neutrino of type 1 follows from the definition~\eqref{eq:adef}.
In the EFT, equation \eqref{eq:adef} translates into the ratio of the above two quantities and reads (including corrections of order $\Delta/M$)
\begin{equation}
\epsilon_{1,{\rm direct}}^{T=0}= \frac{\Gamma^{\ell,T=0}_{11}-\Gamma^{\bar{\ell},T=0}_{11}}{\Gamma^{T=0}_{11}} = 
\left[   (1 -2 \ln 2) + (3 - 4 \ln 2)\frac{\Delta}{M}   \right] \frac{{\rm{Im}}\left[ (F_{1}^{*}F_{2})^2\right]}{8 \pi |F_{1}|^2} .
\label{b14}
\end{equation}
Similarly we may obtain the direct CP asymmetry for the leptonic decay of a neutrino of type 2 just by 
changing $F_1 \leftrightarrow F_2$ and $\Delta \to -\Delta$ in the above formula:
\begin{equation}
\epsilon_{2,{\rm direct}}^{T=0}= -\left[   (1 -2 \ln 2) -(3 - 4 \ln 2)\frac{\Delta}{M}   \right] \frac{{\rm{Im}}\left[ (F_{1}^{*}F_{2})^2\right]}{8 \pi |F_{2}|^2}, 
\label{b15}
\end{equation}
where we have used ${\rm Im}\left[ (F_{2}^{*}F_{1})^2\right] = - {\rm Im}\left[ (F_{1}^{*}F_{2})^2\right]$.
The result agrees with the original result~\cite{Covi:1996wh} and following confirmations, 
like the more recent~\cite{Fong:2013wr}, after accounting for the different definition of the Yukawa couplings\footnote{
Our couplings are the complex conjugate of the couplings in~\cite{Covi:1996wh} and~\cite{Fong:2013wr}.}. 

It is useful to compare equations \eqref{eq15} and \eqref{eq16} with \eqref{b3} and \eqref{b6} respectively. 
It follows that 
\begin{eqnarray}
&& |A|^2 = \frac{M}{16 \pi},\\
&& {\rm{Re}}(B)={\rm{Re}}(C),\\
&& {\rm{Im}}(B)={\rm{Im}}(C)=\frac{1}{16 \pi} \left[   (-1 +2 \ln 2) + (-3 + 4 \ln 2)\frac{\Delta}{M}   \right] \, .
\end{eqnarray}
Replacing the above expressions in \eqref{cpzero} one gets back \eqref{b14}.
The condition ${\rm{Re}}(B)={\rm{Re}}(C)$ requires both ${\rm{Im}}(B)$ and ${\rm Im}\left[ (F_{1}^{*}F_{J})^2\right]$ 
to be different from zero to produce a non-vanishing CP asymmetry.
The first request is at the origin of the condition stated at the end of section~\ref{sec:rev}: 
the relevant two-loop diagrams for the CP asymmetry are those that can be cut with two cuts into three tree-level
diagrams. This guarantees that after a first cut through the lepton (or anti-lepton) propagator 
the remaining one-loop diagram (what is called $B$ above) develops a complex phase.
The second request is fulfilled if there are at least two Majorana neutrino generations with different complex Yukawa couplings. 
In fact only $J = 2$ contributes to the asymmetry in \eqref{eq15} and \eqref{eq16}.

\section{Matching $a_{II}$}
\label{sec:finiteT}
In order to evaluate the leading thermal correction to the direct CP asymmetry, i.e., $\Gamma^{\ell,T}_{II,{\rm direct}}-\Gamma^{\bar{\ell},T}_{II,{\rm direct}}$, 
we need to compute the Wilson coefficients $a_{II}$ of the dimension-five operators in~\eqref{eq:efflag}.
We have seen that at order $F^2$ in the Yukawa couplings the coefficients $a_{II}$ do not contribute to the 
asymmetry, hence, in this section, we will give them at order~$F^4$. 
They also depend linearly on some SM couplings, in particular the four-Higgs and gauge couplings.
The coefficients $a_{II}$ are determined by matching four-point Green's functions with two external Majorana neutrinos and 
two external Higgs bosons computed in the fundamental theory with the corresponding vertices in the EFT. 
In particular, we may consider a Higgs boson with momentum  $q^\alpha \sim T \ll M$ scattering 
off a Majorana neutrino at rest in the reference frame  $v^{\alpha}=(1,\vec{0})$.
In the matching, we integrate out loop momenta of order $M$, hence 
the momentum of the Higgs boson can eventually be set to zero and the matching done in the vacuum.
Thermal corrections do not affect the matching but the CP asymmetry through the Higgs thermal condensate.
Because the Higgs thermal condensate is real, we just need to compute the imaginary parts of $a_{II}$. 
This can be done by using standard cutting rules at $T=0$. 
Diagrams with cuts through lepton propagators contribute to the leptonic component of $a_{II}$, 
$a_{II}^\ell$, while diagrams with cuts through anti-lepton propagators contribute to the anti-leptonic component 
of $a_{II}$, $a_{II}^{\bar{\ell}}$. Not the entire cut diagram contributes to the asymmetry.
The part of the cut diagram that contributes to the asymmetry can be isolated using the same arguments 
developed in the previous section and is proportional to~${\rm Im}\left[ (F_{1}^{*}F_{2})^2\right]$.

The diagrams that enter the matching of ${\rm Im}\,a_{II}^{\ell}$ and ${\rm Im}\,a_{II}^{\bar{\ell}}$
at order $F^4$ and at first order in the SM couplings together with details of the calculation can be found in appendix~\ref{AppendixB}. 
The final result reads up to order $\Delta/M$ (only terms contributing to the asymmetry are displayed):
\begin{eqnarray}
{\rm{Im}} \, a^{\ell}_{11}=-{\rm{Im}} \, a^{\bar{\ell}}_{11} &=&
\frac{{\rm{Im}}\left[ (F_1^{*}F_2)^2\right] }{(16\pi)^2} 
\left\lbrace   6 \lambda \left[ 1+\ln 2-\left( 2-\ln 2 \right)\frac{\Delta}{M}\right]  
\right.
\nonumber\\
&& \hspace{1.1cm}
- \frac{3}{8}g^2 \left[ 2 - \ln2 +\left(3 - 5 \ln 2\right) \frac{\Delta}{M}  \right]  
\nonumber\\
&& \hspace{1.1cm}
\left.
- \frac{g'^2}{8}\left[ 4 - \ln2 +\left(1 - 5 \ln 2\right) \frac{\Delta}{M}  \right]  \right\rbrace ,
\label{match1} \\
{\rm{Im}} \, a^{\ell}_{22}=-{\rm{Im}} \, a^{\bar{\ell}}_{22} &=& 
-\frac{{\rm{Im}}\left[ (F_1^{*}F_2)^2\right] }{(16\pi)^2} 
\left\lbrace   6 \lambda \left[ 1+\ln 2+\left( 2-\ln 2 \right)\frac{\Delta}{M}\right]  
\right.
\nonumber\\
&& \hspace{1.2cm}
- \frac{3}{8}g^2\left[ 2 - \ln2 -\left(3 - 5 \ln 2\right) \frac{\Delta}{M}  \right]  
\nonumber\\
&& \hspace{1.1cm}
\left.
- \frac{g'^2}{8}\left[ 4 - \ln2 -\left(1 - 5 \ln 2\right) \frac{\Delta}{M}  \right]  \right\rbrace ,
\label{match2}
\end{eqnarray}
where $\lambda$ is the four-Higgs coupling, and $g$ and $g'$ are the SU(2)$_L$ and U(1)$_Y$ gauge couplings respectively.
Note the sign difference between ${\rm{Im}} \, a^{\ell}_{II}$ and ${\rm{Im}} \, a^{\bar{\ell}}_{II}$.
We remark that at this order the result does not depend on the top-Yukawa coupling, $\lambda_t$.

\section{Thermal corrections to the direct asymmetry}
\label{sec:direct}
We may now proceed to calculate the thermal corrections to the widths and CP asymmetries
of the two Majorana neutrinos, assuming that the thermal bath of SM particles is at rest with respect to the
Majorana neutrinos and the reference frame. It is convenient to split both the neutrino width, 
$\Gamma_{II}=\Gamma_{II}^{T=0}+ \Gamma_{II}^{T}$, and the CP asymmetry, 
$\epsilon_{I}=\epsilon^{T=0}_{I} + \epsilon_{I}^{T}$, into a zero temperature and a thermal part.

\begin{figure}[htb]
\centering
\includegraphics[scale=0.6]{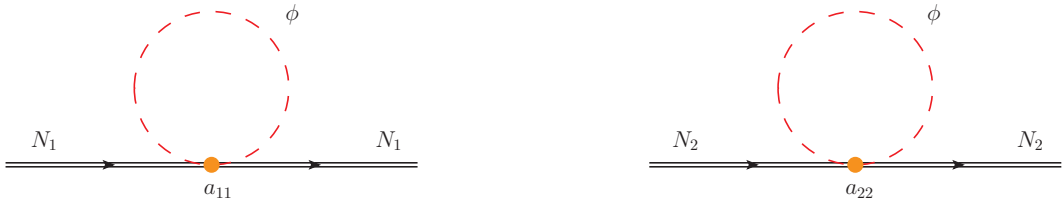}
\caption{Tadpole diagrams responsible for the leading thermal corrections to the neutrino widths 
and CP asymmetries in the EFT. We show in red particles belonging to the thermal bath whose momentum is of order $T$.}
\label{fig:tadpoles} 
\end{figure}

\subsection{Neutrino of type 1}
\label{sec:direct1}
We consider first neutrinos of type 1, which are assumed to be lighter than those of type~2.
The zero-temperature width at leading order has been written in \eqref{Gamma1T0}.
The leading thermal correction to the width has been calculated in~\cite{Salvio:2011sf,Laine:2011pq,Biondini:2013xua}
and can be easily re-derived from \eqref{asysec3}, \eqref{higgscondensate} and \eqref{agen}.
The expression of the width up to order  $F^2 \lambda \times (T/M)^2$ reads  
\begin{equation}
\Gamma_{11} = \Gamma_{11}^{T=0}+\Gamma_{11}^{T} =\frac{M}{8\pi} |F_1|^2 \left[ 1-\lambda\left(\frac{T}{M}\right)^2 \right]  .
\label{deno}
\end{equation}

The in-vacuum part of the direct CP asymmetry, $\epsilon^{T=0}_{1,{\rm direct}}$, can be read off~\eqref{b14}. 
In order to obtain $\epsilon^{T}_{1,{\rm direct}}$, one has to evaluate $\Gamma^{\ell,T}_{11,{\rm direct}}-\Gamma^{\bar{\ell},T}_{11,{\rm direct}}$.
Thermal corrections are encoded into the Higgs thermal condensate represented by the first tadpole diagram shown in figure~\ref{fig:tadpoles}.
From  \eqref{asysec3}, \eqref{higgscondensate} and \eqref{match1} it follows
\begin{eqnarray}
\Gamma^{\ell,T}_{11,{\rm direct}}-\Gamma^{\bar{\ell},T}_{11,{\rm direct}} &=& 
\frac{{\rm{Im}}\left[ (F^{*}_1 F_2)^2\right] }{64 \pi^2}  
\left\lbrace   \lambda \left[ 1 + \ln 2-\left( 2-\ln 2 \right)\frac{\Delta}{M}\right]  
\right.
\nonumber\\
&& \hspace{-4cm}
\left.
- \frac{g^2}{16}\left[ 2- \ln2 +\left( 3 - 5 \ln 2\right) \frac{\Delta}{M}  \right]  
- \frac{g'^2}{48}\left[ 4- \ln2 +\left( 1 - 5 \ln 2\right) \frac{\Delta}{M}  \right]  \right\rbrace  \frac{T^2}{M}.
\label{gammaphi} 
\end{eqnarray} 
From \eqref{asysec2bis}, \eqref{EQ17}, \eqref{deno} and \eqref{gammaphi}, and considering that 
$\Delta\Gamma_{1,{\rm direct}}^{\rm mixing} =0$, we obtain then the thermal part of the CP asymmetry 
generated from the decay of Majorana neutrinos of type 1 at leading order in the SM couplings, at order $T^2/M^2$ and at order $\Delta/M$:
\begin{eqnarray}
\epsilon^{T}_{1,{\rm direct}} &=&\frac{{\rm{Im}}\left[ (F^{*}_1 F_2)^2\right] }{8 \pi |F_{1}|^2}  \left(  \frac{T}{M} \right)^2 
\left\lbrace   \lambda \left[ 2-\ln 2+\left( 1-3\ln 2 \right) \frac{\Delta}{M}\right]  
\right.
\nonumber \\
&& 
\left.
- \frac{g^2}{16}\left[ 2- \ln2 +\left( 3 - 5 \ln 2\right) \frac{\Delta}{M}  \right]  
- \frac{g'^2}{48}\left[ 4- \ln2 +\left( 1 - 5 \ln 2\right) \frac{\Delta}{M}  \right]  \right\rbrace.
\label{CPnu1}
\end{eqnarray}

\subsection{Neutrino of type 2}
\label{sec:direct2}
The in-vacuum contribution to the CP asymmetry of Majorana neutrinos of type~2 can be read off~\eqref{b15}.
Thermal contributions of the type \eqref{asysec3}, can be computed as for neutrinos of type 1,  
the relevant diagram being the second diagram of figure~\ref{fig:tadpoles}. 
They may be read off \eqref{gammaphi} and \eqref{CPnu1} after the replacements 
$F_1 \leftrightarrow F_2$, $M \to M_2$ and $\Delta \to -\Delta$.

\begin{figure}[hbt]
\centering
\includegraphics[scale=0.52]{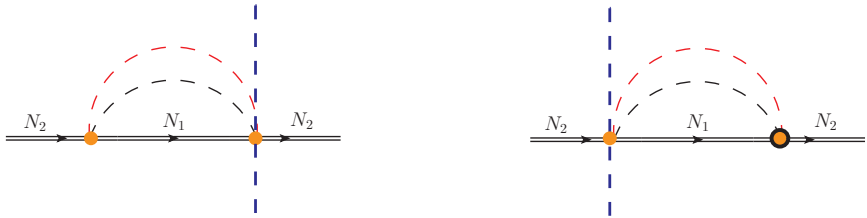}
\caption{Diagrams contributing in the EFT to the CP asymmetry of the Majorana neutrino of type 2 (see text). 
The orange dot stands for the vertex $-i {\rm Re}\,(F^{*}_1 F_2)/M$; the circled dot has opposite sign.
The dot with a cut selects the leptonic (or anti-leptonic) decay components: 
$-3(F_1 F^{*}_2)\lambda/(8\pi M)$ (or $-3(F_2 F^{*}_1)\lambda/(8\pi M)$) for incoming neutrino of type 1. 
Propagators on the right of the cut are complex conjugate.
Red dashed lines indicate thermal Higgs bosons, while black dashed lines indicate 
Higgs bosons carrying a momentum and energy of order~$\Delta$. 
}
\label{fig:DeltaEFT}
\end{figure}

If the neutrino of type 2 is heavier than the neutrino of type 1, there may be an additional source of CP asymmetry coming from 
diagrams where, after the cut through the lepton (or anti-lepton), the remaining one-loop subdiagram develops 
an imaginary part because of the kinematically allowed transition $\nu_{R,2}\to\nu_{R,1} + $ Higgs boson.
Such a transition involves a momentum transfer of order $\Delta$. Since $\Delta \ll M$, terms coming from momentum 
regions of order $\Delta$ have been excluded from the matching and do not contribute to $a_{IJ}$.
However, they do contribute in the EFT. 

\begin{figure}[ht]
\centering
\includegraphics[scale=0.52]{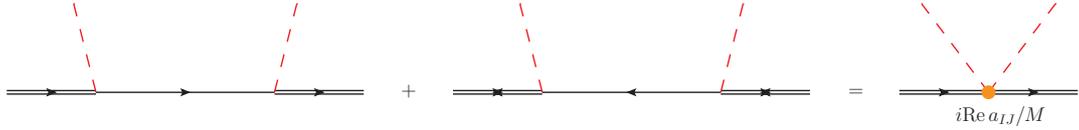}
\caption{On the left-hand side the diagrams in the fundamental theory that match the real part of $a_{IJ}$ at order $F^2$ (right-hand side).
Red dashed lines indicate external Higgs bosons with a soft momentum much smaller than the mass of the Majorana neutrinos.}
\label{fig:treeMatch} 
\end{figure}

The leading order diagrams in the EFT are shown in figure~\ref{fig:DeltaEFT}.\footnote{
The corresponding diagrams in the full theory are diagrams 1)-6) in figure~\ref{fig:new_Higgs_direct}.} 
They may be understood as the mixing of two dimension five operators in the EFT, 
hence they contribute to the direct CP asymmetry \eqref{asysec2bis} through the term $\Delta\Gamma_{2,{\rm direct}}^{\rm mixing}$.
At our accuracy, for the uncut vertex, we just need to consider the real parts of the dimension five operators 
mixing neutrinos of type 1 with neutrinos of type 2. The corresponding vertex, 
shown with an orange dot in figure~\ref{fig:DeltaEFT}, is $i {\rm Re}\,a_{12}/M$. 
The real part of $a_{IJ}$ can be computed at order $F^2$ by matching the two tree-level diagrams 
shown in the left-hand side of figure~\ref{fig:treeMatch} with the corresponding vertex in the EFT. 
The result reads
\begin{equation}
{\rm{Re}}\, a_{IJ}=-\frac{F_IF_J^*+F_I^*F_J}{2}.
\label{re}
\end{equation}
The contribution from the cut is $-2 \times 1/M \times (3\,F_{I}^{*}F_{J}\lambda/(16\pi))$ for the leptonic cut 
and $-2 \times 1/M \times (3\,F_{J}^{*}F_{I}\lambda/(16\pi))$ for the anti-leptonic one, where $I$ is the outgoing neutrino and $J$ the ingoing one.

The momentum flowing in the diagrams of figure~\ref{fig:DeltaEFT} can be of order $T$ or of order $\Delta$. 
If the momentum flowing in both loops is of order $T$ this contributes to the 
asymmetry $\Gamma^{\ell,T}_{22,{\rm direct}}-\Gamma^{\bar{\ell},T}_{22,{\rm direct}}$ at order $T^3/M^2$; 
if the momentum flowing in both loops is of order $\Delta$ this contributes to the 
asymmetry at order $\Delta^3/M^2$. Both contributions are beyond our accuracy.
If instead one Higgs boson carries a momentum and energy of order $T$ and the other a momentum and energy of order $\Delta$, 
then this momentum region contributes to the asymmetry at order $T^2\Delta /M^2$, which is inside our accuracy.
The color code used for the Higgs bosons in figure~\ref{fig:DeltaEFT} identifies this specific momentum region. 
Its contribution to the direct asymmetry of Majorana neutrinos of type 2 is
\begin{equation}
\Delta\Gamma_{2,{\rm direct}}^{\rm mixing} = \frac{{\rm{Im}}\left[ (F^{*}_1 F_2)^2\right] }{16 \pi^2} \lambda \frac{T^2\Delta}{M^2}.
\label{gammaDelta}
\end{equation}
Summing this to the CP asymmetry of the Majorana neutrino of type 2 obtained from the tadpole diagram 
of figure~\ref{fig:tadpoles}, and discussed at the beginning of this section, we obtain that 
the thermal correction to the direct CP asymmetry of the Majorana neutrino of type 2 
at leading order in the SM couplings, at order $T^2/M^2$ and at order $\Delta/M$ is
\begin{eqnarray}
\epsilon^{T}_{2,{\rm direct}} &=& -\frac{{\rm{Im}}\left[ (F^{*}_1 F_2)^2\right] }{8 \pi |F_{2}|^2}  \left(  \frac{T}{M} \right)^2 
\left\lbrace   \lambda \left[ 2-\ln 2-\left( 9 - 5\ln 2 \right) \frac{\Delta}{M}\right]  \right.
\nonumber \\
&&
\left.
- \frac{g^2}{16}\left[ 2- \ln2 - 7 \left( 1 - \ln 2\right) \frac{\Delta}{M}  \right]  
- \frac{g'^2}{48}\left[ 4- \ln2 -\left( 9 - 7 \ln 2\right) \frac{\Delta}{M}  \right]  \right\rbrace.
\label{CPnu2}
\end{eqnarray}

We observe that in the exact degenerate limit ($\Delta \to 0$), the single direct CP asymmetries $\epsilon_{1,{\rm direct}}$
and $\epsilon_{2,{\rm direct}}$ do not vanish. However, the sum of \eqref{EQ17} with \eqref{gammaphi}, and 
with the corresponding expressions for the type 2 neutrino does vanish. 
This sum is the CP-violating parameter defined in~\cite{Pilaftsis:1998pd}.

\section{Indirect asymmetry}
\label{sec:indirect}
The indirect CP asymmetry is made of all contributions that exhibit the phenomenon of resonant enhancement (see section~\ref{sec:rev}).
It may be understood as originating from the mixing between the different neutrino species 
that makes the mass eigenstates different from the CP eigenstates~\cite{Flanz:1996fb}.
This mixing is described by the EFT. 
In the following we will compute the indirect CP asymmetry at leading order and its first thermal correction. 
Besides the hierarchies $M\gg T \gg M_{W}$ and $M \gg \Delta$ we will not assume any special relation between 
$\Delta$ and the neutrino decay widths. In particular we will allow for the resonant case 
$\Delta \sim \Gamma_{11},\Gamma_{22}$ and resum the widths in the neutrino propagators.
Nevertheless we will treat the mixing perturbatively, which amounts at requiring
$\Delta^2 + (\Gamma_{22}-\Gamma_{11})^2/4 \gg M^2 \, [{\rm Re}(F_1^*F_2)]^2/(16\pi)^2$ 
(this condition can be inferred from the right-hand side of the following equation~\eqref{gammalt0indirect}; 
see also~\cite{Garny:2011hg}).\footnote{
Relaxing this condition does not pose conceptual problems. A non-perturbative mixing will affect, however, 
both the direct and indirect CP asymmetries and make their analytical expressions less compact.
For the indirect asymmetry, this has been considered without resummation of the widths in~\cite{Flanz:1996fb}.}  

Mixing between the different neutrino generations in the effective Lagrangian \eqref{eq:efflag} 
is induced by the off-diagonal elements of $\Gamma_{IJ}^{T=0}$, 
\begin{equation}
\Gamma_{IJ}^{T=0} = \frac{M}{16\pi}\left( F_I^*F_J + F_J^*F_I \right),
\label{gammamixing}
\end{equation}
which can be obtained from the absorbtive part of diagram $1)$ in figure~\ref{Fig2} and the corresponding 
one with an anti-lepton in the loop~\cite{Flanz:1996fb,Buchmuller:1997yu}
(for $I=J=1$ \eqref{gammamixing} gives back \eqref{Gamma1T0}), and by the off-diagonal elements of $a_{IJ}$.
The imaginary part of $a_{IJ}$  is 
\begin{equation}
{\rm Im}\, a_{IJ} = -\frac{3}{16\pi}(F_{J}F_{I}^{*}+F_{I}F_{J}^{*})\lambda.
\label{agenbis}
\end{equation}
The real part of $a_{IJ}$ has been computed at order $F^2$ in the previous section and can be read off~\eqref{re}. 

\begin{figure}[ht]
\centering
\includegraphics[scale=0.52]{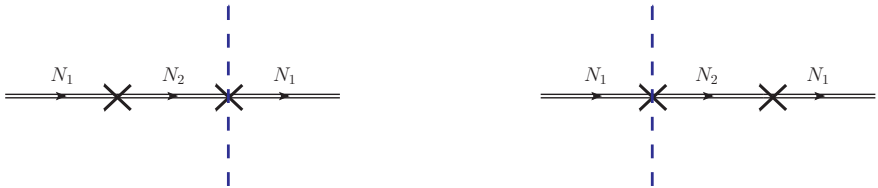}
\caption{Diagrams showing in the EFT a neutrino of type 1 decaying into a lepton after mixing with a neutrino of type 2.
The cross stands for the mixing vertex $-\Gamma_{IJ}^{T=0}/2$. 
The cross with a cut selects the leptonic (or anti-leptonic) decay components: 
$M(F_I^*F_J)/(16\pi)$ (or $M(F_J^*F_I)/(16\pi)$).
Propagators on the right of the cut are complex conjugate.
Because the mixing vertex is real, circled and uncircled vertices coincide~\cite{Denner:2014zga}.
}
\label{fig:indirectEFT}
\end{figure}

At zero temperature and at order $F^4$ the width of a neutrino of type 1 that decays into a lepton after mixing with a neutrino of type 2 
is given in the EFT by the sum of the cuts on the diagrams shown in figure~\ref{fig:indirectEFT}.
The diagrams are amputated of the external legs and evaluated at the pole of the propagator of the (incoming and outgoing) neutrino of type 1.
If the width is of the order of $\Delta$, then it should be resummed so that the (complex) pole of the neutrino of type 1 is at $-i\Gamma_{11}^{T=0}/2$
and the pole of the intermediate neutrino of type 2 is at  $\Delta -i\Gamma_{22}^{T=0}/2$.
The crossed vertex in figure~\ref{fig:indirectEFT} stands for the mixing vertex $-\Gamma_{IJ}^{T=0}/2$, 
where $I$ identifies the outgoing and $J$ the incoming neutrino. 
The cut through the vertex selects the decay into a lepton or an anti-lepton.
In the first case, the value of the cut is $M(F_I^*F_J)/(16\pi)$, in the second case it is $M(F_J^*F_I)/(16\pi)$.
For leptonic cuts the diagrams in figure~\ref{fig:indirectEFT} give
\begin{equation}
\Gamma^{\ell,T=0}_{11,{\rm indirect}} = \frac{M}{16\pi} F_1^*F_2\frac{i}{-\Delta+i(\Gamma_{22}^{T=0}-\Gamma_{11}^{T=0})/2}
\left(-\frac{M}{16\pi}\right)\frac{F_1^*F_2+F_2^*F_1}{2} + {\rm c.c.},
\label{gammalt0indirect}
\end{equation}
where c.c. stands for complex conjugate. For anti-leptonic cuts the diagrams in figure~\ref{fig:indirectEFT}  give 
$\Gamma^{\bar{\ell},T=0}_{11,{\rm indirect}}$, which is the same as \eqref{gammalt0indirect} but with the change $F_1^*F_2 \leftrightarrow F_2^*F_1$ in the mixing vertices.
The indirect CP asymmetry at $T=0$ for a Majorana neutrino of type 1 is then
\begin{equation}
\epsilon_{1,{\rm indirect}}^{T=0}= \frac{\Gamma^{\ell,T=0}_{11,{\rm indirect}}-\Gamma^{\bar{\ell},T=0}_{11,{\rm indirect}}}{\Gamma^{T=0}_{11}} 
= - \frac{{\rm{Im}}\left[ (F^{*}_1 F_2)^2\right] }{16 \pi |F_{1}|^2}  \frac{M \, \Delta}{\Delta^2 + (\Gamma_{22}^{T=0}-\Gamma_{11}^{T=0})^2/4}.
\label{indirect1T0}
\end{equation}
Similarly one obtains the indirect CP asymmetry at $T=0$ for a Majorana neutrino of type~2
\begin{equation}
\epsilon_{2,{\rm indirect}}^{T=0}= \frac{\Gamma^{\ell,T=0}_{22,{\rm indirect}}-\Gamma^{\bar{\ell},T=0}_{22,{\rm indirect}}}{\Gamma^{T=0}_{22}} 
= - \frac{{\rm{Im}}\left[ (F^{*}_1 F_2)^2\right] }{16 \pi |F_{2}|^2}  \frac{M \, \Delta}{\Delta^2 + (\Gamma_{22}^{T=0}-\Gamma_{11}^{T=0})^2/4}.
\label{indirect2T0}
\end{equation}
We recall that $\Gamma_{II}^{T=0} = M|F_I|^2/(8\pi)$.

The above result for the indirect asymmetry at $T=0$ agrees with~\cite{Buchmuller:1997yu} (see also~\cite{Garny:2011hg} and discussion therein). 
It agrees with~\cite{Anisimov:2005hr} by remarking that the additional term proportional to $\log(M^2_2/M^2_1)$ there 
is a contribution of relative order $F^6$ to the CP asymmetry and therefore beyond our accuracy.
Whenever we can neglect the width $\Gamma_{11}^{T=0}$, equations~\eqref{indirect1T0} and \eqref{indirect2T0}
agree with~\cite{Pilaftsis:1997dr,Pilaftsis:1997jf,Pilaftsis:1998pd,Pilaftsis:2003gt,Dev:2014laa}.
Finally, we notice that in the framework of the Kadanoff--Baym evolution equations 
(see for instance \cite{Garny:2011hg,Frossard:2012pc,Garbrecht:2014aga}) the quantity related to the CP asymmetry 
is a modification of the above one that accounts for coherent transitions between the Majorana neutrino mass eigenstates.

The computation done above shows that, although at $T=0$ there should be in general no advantage in using the EFT, there is some in computing the indirect CP asymmetry.
In fact, the EFT naturally separates the physics of the Majorana neutrino decay, which goes into the widths and the mixing vertices, 
from the quantum-mechanical physics of the neutrino oscillations. 
This separation is well depicted in the Feynman diagrams of figure~\ref{fig:indirectEFT}.
It also makes more apparent the potentially resonant behaviour of the contribution.

Thermal corrections to \eqref{gammalt0indirect} affect masses, widths and mixing vertices. 
From \eqref{asysec3} (generalized to off-diagonal elements), \eqref{higgscondensate} and 
\eqref{agenbis} it follows that the leading thermal correction to the width matrix is of relative size $\lambda T^2/M^2$:
\begin{equation}
\Gamma^T_{IJ}=-\frac{\lambda T^2}{16\pi M}(F_IF_J^*+F_I^*F_J).
\label{gammat}
\end{equation}
The thermal correction to the mass matrix follows from \eqref{re} and \eqref{higgscondensate}, and is of relative size $T^2/M^2$:
\begin{equation}
M^T_{IJ}=\frac{T^2}{12 M}(F_IF_J^*+F_I^*F_J).
\label{mass}
\end{equation}
The mass thermal correction \eqref{mass} differs from the one used in~\cite{Pilaftsis:1997jf} and taken from~\cite{Weldon:1982bn}. 
The reason for the difference is that the thermal correction computed in~\cite{Weldon:1982bn}
refers to a massless neutrino while the one written above refers to a neutrino in the heavy mass limit.
In the massless case the neutrino gets a thermal mass both from fermions and bosons in the medium, whereas 
in the heavy-mass case, as can be immediately read off the effective Lagrangian \eqref{eq:efflag},
fermion contributions are suppressed in $T/M$ and only Higgs bosons contribute.

If we restrict to the leading corrections, we may neglect the thermal correction to the 
decay matrix, which is suppressed by $\lambda$, and keep only the thermal correction to the mass matrix. 
This modifies the mixing vertex in figure~\ref{fig:indirectEFT} from 
 $-\Gamma_{IJ}^{T=0}/2$ to  $-\Gamma_{IJ}^{T=0}/2 -i M^T_{IJ}$ and the mass $\Delta$ in the 
intermediate propagator to $\Delta + M^T_{22}-M^T_{11}$.
If we neglect corrections of relative order $\lambda$, cuts are not affected by thermal effects, so that 
\begin{eqnarray}
\Gamma^{\ell,T}_{11,{\rm indirect}} &=&\bigg[ \frac{M}{16\pi} F_1^*F_2\frac{i}{-\Delta
- (|F_2|^2-|F_1|^2)T^2/(6M) +i(\Gamma_{22}^{T=0}-\Gamma_{11}^{T=0})/2}
\nonumber\\
&&
\hspace{1.5cm}
\times\left(-\frac{M}{16\pi} -i \frac{T^2}{6M}\right)\frac{F_1^*F_2+F_2^*F_1}{2} + {\rm c.c.}\bigg]
- \Gamma^{\ell,T=0}_{11,{\rm indirect}} \;,
\label{gammaltindirect}
\end{eqnarray}
which is valid at leading order in $T/M$.
Similarly $\Gamma^{\bar{\ell},T}_{11,{\rm indirect}}$ is given by \eqref{gammaltindirect} but with the change $F_1^*F_2 \leftrightarrow F_2^*F_1$ in the mixing vertices.
The leading thermal correction to the indirect CP asymmetry for a Majorana neutrino of type 1 is then 
\begin{equation}
\epsilon_{1,{\rm indirect}}^{T}= -\frac{\epsilon_{1,{\rm indirect}}^{T=0}}{3} \,\left(|F_2|^2-|F_1|^2\right)\,
\frac{M\Delta}{\Delta^2 + (\Gamma_{22}^{T=0}-\Gamma_{11}^{T=0})^2/4}\,\frac{T^2}{M^2},
\label{indirect1T}
\end{equation}
and analogously the thermal correction to the indirect CP asymmetry for a neutrino of type~2 is 
\begin{equation}
\epsilon_{2,{\rm indirect}}^{T}= -\frac{\epsilon_{2,{\rm indirect}}^{T=0}}{3} \,\left(|F_2|^2-|F_1|^2\right)\,
\frac{M\Delta}{\Delta^2 + (\Gamma_{22}^{T=0}-\Gamma_{11}^{T=0})^2/4}\,\frac{T^2}{M^2}.
\label{indirect2T}
\end{equation}
Note that the indirect asymmetry vanishes for each neutrino type in the exact degenerate limit 
$\Delta \to 0$~\cite{Buchmuller:1997yu,Pilaftsis:1998pd}.

\section{Flavour and CP asymmetry}
\label{sec:flavour}
In the previous sections we have computed the CP asymmetry, both direct and indirect, in the so-called unflavoured approximation, 
i.e., we have computed the CP parameter, defined in \eqref{eq:adef}, as a sum over the different lepton flavours.
This is the relevant CP asymmetry parameter when the flavour composition of the quantum states of the leptons (anti-leptons) in the thermal plasma 
has no influence on the final lepton asymmetry.  If this is not the case, then one has to define a CP asymmetry for each lepton family. 
The unflavoured regime is found to be an appropriate choice at high temperatures, namely $T \simg 10^{12}$ GeV,  
while the different lepton flavours are resolved at lower temperatures~\cite{Nardi:2005hs, Nardi:2006fx}. 
In~\cite{Campbell:1992jd, Cline:1993bd} it was shown how to estimate the temperature at which the different lepton flavours are resolved 
considering the interactions induced by charged lepton Yukawa couplings in the most general seesaw type-I Lagrangian 
(we have not included these interactions in the Lagrangian \eqref{eq3}; one can find them, e.g., in~\cite{Davidson:2008bu}). 
It is found that at $T \approx 10^{12}$ GeV, the interaction rates involving the $\tau$-doublet are faster than the universe expansion rate. 
Hence the $\tau$-flavour is resolved by the thermal bath, while the $e$- and $\mu$-flavours remain unresolved. 
At temperatures of about $10^9$ GeV all three flavours are resolved from the charged Yukawa coupling interactions.
The importance of flavour effects in leptogenesis has been investigated in the literature in many different directions, see, e.g.,~\cite{Blanchet:2006be, DeSimone:2006nrs}.

In order to investigate how the flavour affects our approach, we start with the definition of the CP asymmetry, $\epsilon_{fI}$, 
generated by the $I$-th heavy neutrino decaying into leptons and anti-leptons of flavour $f$: 
\begin{equation}
\epsilon_{fI}=
\frac{ \Gamma(\nu_{R,I} \to \ell_{f} + X)-\Gamma(\nu_{R,I} \to \bar{\ell}_{f}+ X )  }
{\sum_f \Gamma(\nu_{R,I} \to \ell_{f} + X ) + \Gamma(\nu_{R,I} \to \bar{\ell}_{f}+ X)} \, .
\label{eq:adef_fla}
\end{equation}
The difference with respect to  \eqref{eq:adef} is that we do not sum over the flavour index $f$ in the numerator.
Following the same order adopted for the unflavoured case, we will, 
first, compute the flavoured direct and indirect CP asymmetries at $T=0$, and then the CP asymmetries at finite temperature.

It is straightforward to extend the derivation of section~\ref{sec:zeroT} for the direct CP asymmetry at $T=0$ in the unflavoured case 
to the CP asymmetry in the flavoured case.
In the flavoured case one has simply to omit the sum over the flavour index $f$ in \eqref{b3} and \eqref{b6},  
obtaining for the CP asymmetry in the neutrino of type~1 decays
\begin{eqnarray}
&& \epsilon_{f1} =   
\nonumber\\
&& \sum_J \frac{\left( {\rm{Re}}(B)-{\rm{Re}}(C) \right) {\rm{Re}} \left[ (F_{1}^{*}F_{J})(F_{f1}^*F_{fJ})\right] 
- \left( {\rm{Im}}(B)+{\rm{Im}}(C) \right) {\rm{Im}} \left[ (F_{1}^{*}F_{J})(F_{f1}^*F_{fJ}) \right] }{|F_{1}|^2} .
\nonumber\\
\label{cpzero_fla}
\end{eqnarray}
The calculation of the diagrams in figure~\ref{Fig2} leads to the same results for the functions $A$, $B$ and $C$: 
the loop calculation is unaffected by the different treatment of the flavour. 
Note that additional two-loop diagrams, similar to 2) and 3) of figure~\ref{Fig2} but involving only lepton (or anti-lepton) internal lines,  
are not allowed by the Feynman rules of \eqref{eq3}.
Therefore the direct CP asymmetry at $T=0$ for the neutrino of type~1 decay into leptons of flavour $f$ 
reads up to order $\Delta /M$ 
\begin{equation}
\epsilon_{f1,{\rm direct}}^{T=0} = 
\left[   (1 -2 \ln 2) + (3 - 4 \ln 2)\frac{\Delta}{M}   \right] \frac{{\rm{Im}}\left[ (F_{1}^{*}F_{2})(F_{f1}^*F_{f2}) \right]}{8 \pi |F_{1}|^2} .
\label{CPdir_1fla}
\end{equation}
The result for $\epsilon_{f2,{\rm direct}}^{T=0}$ can be obtained from the above formula by changing $F_1 \leftrightarrow F_2$ and $\Delta \to -\Delta$.
The results agree in the nearly degenerate limit with the flavoured CP asymmetry obtained in~\cite{Fong:2013wr}. 

\begin{figure}[ht]
\centering
\includegraphics[scale=0.57]{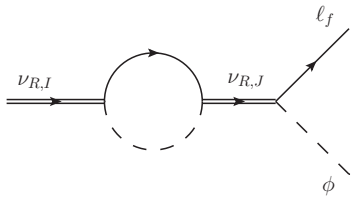}
\caption{One-loop self-energy diagram responsible for an additional contribution to the indirect CP asymmetry in the flavoured case. 
Note that only heavy-neutrino propagators with forward arrow appear, namely $\langle 0| T(\psi \bar{\psi}) |0 \rangle$. }
\label{fig:flavor_1} 
\end{figure}

\begin{figure}[ht]
\centering
\includegraphics[scale=0.53]{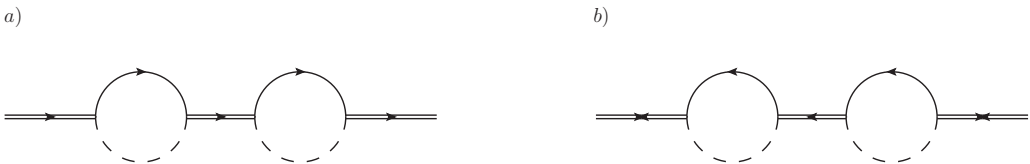}
\caption{Two-loop self-energy diagrams in the fundamental theory contributing to the indirect CP asymmetry at $T=0$ in the flavoured case only. 
Diagram $a)$ admits two cuts through lepton lines, whereas diagram $b)$ admits two cuts through anti-lepton lines.}
\label{fig:flavor_2} 
\end{figure}

We can compute the flavoured indirect CP asymmetry at $T=0$ either in the fundamental or in the effective theory.
In the fundamental theory, besides the diagrams that appear in the unflavoured case,  
one has to consider also the interference between the tree-level diagram of figure~\ref{drind} 
with the additional one-loop diagram shown in figure~\ref{fig:flavor_1}. 
This contribution is equivalent to cutting through lepton or anti-lepton lines respectively the two-loop diagrams $a)$ and $b)$ shown in figure~\ref{fig:flavor_2}. 
The additional diagrams give a contribution to the CP asymmetry that is proportional to ${\rm Im}\left[(F_1 F^*_2)  (F^*_{f1} F_{f2})\right]$.
Clearly this contribution vanishes if summed over all flavours $f$. 
For this reason it has not been considered in the unflavoured case.

As argued in section~\ref{sec:indirect}, it is particularly convenient to compute the indirect CP asymmetry in the EFT. 
In fact, the relevant diagrams are the same computed in the unflavoured case, i.e., those shown in figure~\ref{fig:indirectEFT}. 
They already comprise the two additional diagrams of figure~\ref{fig:flavor_2}, 
the only difference being that now the cut through the mixing vertex selects the decay into a specific leptonic (or anti-leptonic) flavour family. 
More specifically the cut stands for $M(F_{fI}^*F_{fJ})/(16\pi)$ (or $M(F_{fJ}^*F_{fI})/(16\pi)$), 
where $I$ is the type of the outgoing and $J$ the type of the incoming neutrino.
Hence the result for the leptonic width of a neutrino of type 1 decaying into a lepton of flavour $f$
can be read off \eqref{gammalt0indirect} by refraining of summing over the flavours in the leptonic cuts
\begin{equation}
\Gamma^{\ell,T=0}_{f11,{\rm indirect}} = \frac{M}{16\pi} F_{f1}^*F_{f2}\frac{i}{-\Delta+i(\Gamma_{22}^{T=0}-\Gamma_{11}^{T=0})/2}
\left(-\frac{M}{16\pi}\right)\frac{F_1^*F_2+F_2^*F_1}{2} + {\rm c.c.}\;.
\label{gammalt0indirect_fla}
\end{equation}
For anti-leptonic cuts the diagrams in figure~\ref{fig:indirectEFT} give the anti-leptonic width, $\Gamma^{\bar{\ell},T=0}_{f11,{\rm indirect}}$, 
which is the same as \eqref{gammalt0indirect_fla} but with the change $F_{f1}^*F_{f2} \leftrightarrow F_{f2}^*F_{f1}$ in the mixing vertices.
The flavoured indirect CP asymmetry at $T=0$ for a Majorana neutrino of type~1 then is\footnote{
A more compact expression follows from 
${\rm{Im}}\left[(F^*_1 F_2)  (F^*_{f1} F_{f2})\right] + {\rm{Im}}\left[(F_1 F^*_2)  (F^*_{f1} F_{f2})\right] = 
2 \,{\rm{Re}}\left[(F^*_1 F_2) \right] \, {\rm{Im}}\left[(F^*_{f1} F_{f2})\right]$.}
\begin{eqnarray}
\epsilon_{f1,{\rm indirect}}^{T=0}=
&-& \frac{{\rm{Im}}\left[(F^*_1 F_2)  (F^*_{f1} F_{f2})\right] }{16 \pi |F_{1}|^2}  \frac{M \, \Delta}{\Delta^2 + (\Gamma_{22}^{T=0}-\Gamma_{11}^{T=0})^2/4}
\nonumber \\ 
&-& \frac{{\rm{Im}}\left[(F_1 F^*_2)  (F^*_{f1} F_{f2})\right] }{16 \pi |F_{1}|^2}  \frac{M \, \Delta}{\Delta^2 + (\Gamma_{22}^{T=0}-\Gamma_{11}^{T=0})^2/4}.
\label{indirect1T0_fla}
\end{eqnarray}
The first line, if summed over all flavours, gives back \eqref{indirect1T0}.
The second line is specific of the flavoured CP asymmetry and would vanish if summed over all flavours, 
indeed, $\sum_f {\rm{Im}}\left[ (F_1 F^*_2) (F^*_{f1} F_{f2})\right] = {\rm{Im}}\left[ |(F_1 F^*_2)|^2 \right]=0$. 
A similar calculation leads to the expression for the flavoured indirect CP asymmetry at $T=0$ for a Majorana neutrino of type~2, 
which follows from \eqref{indirect1T0_fla} after the changes $F_1 \leftrightarrow F_2$ and $\Delta \to -\Delta$:
\begin{eqnarray}
\epsilon_{f2,{\rm indirect}}^{T=0}=
&-& \frac{{\rm{Im}}\left[(F^*_1 F_2)  (F^*_{f1} F_{f2})\right] }{16 \pi |F_{2}|^2}  \frac{M \, \Delta}{\Delta^2 + (\Gamma_{22}^{T=0}-\Gamma_{11}^{T=0})^2/4}
\nonumber \\ 
&-& \frac{{\rm{Im}}\left[(F_1 F^*_2)  (F^*_{f1} F_{f2})\right] }{16 \pi |F_{2}|^2}  \frac{M \, \Delta}{\Delta^2 + (\Gamma_{22}^{T=0}-\Gamma_{11}^{T=0})^2/4}.
\label{indirect2T0_fla}
\end{eqnarray}
The expressions for $\epsilon_{f1,{\rm indirect}}^{T=0}$ and $\epsilon_{f2,{\rm indirect}}^{T=0}$ agree with those that can be found in~\cite{Fong:2013wr} 
when taking the nearly degenerate limit and resumming the widths of both types of neutrino in the heavy-neutrino propagators. 

We conclude by computing the flavoured CP asymmetries at finite temperature. 
Concerning the direct asymmetry, we may identify two type of contributions. 
First, there are contributions coming from the same diagrams considered for the unflavoured case. 
These diagrams contribute also to the flavoured CP asymmetry if the final lepton (or anti-lepton) flavour is resolved.
This amounts at replacing 
\begin{equation}
{\rm{Im}}\left[ (F_1^* F_2)^2\right] \to {\rm{Im}}\left[ (F^{*}_{1}F_{2})(F^{*}_{f1}F_{f2})\right]   \, ,
\label{replacement}
\end{equation}
in the expressions of the Feynman diagrams given in the appendices~\ref{appHiggs} and~\ref{appgauge}.

A second type of contributions comes from diagrams involving only lepton (or anti-lepton) lines.
They would potentially give rise to a CP asymmetry that is proportional to ${\rm{Im}}\left[(F_1F_2^{*})(F^*_{f1}F_{f2})\right]$ 
and that would vanish in the unflavoured case. 
We have examined these diagrams in appendix~\ref{AppendixC} and found that they do not contribute.
Hence, the complete contribution to the matching coefficients ${\rm{Im}} \, a^{\ell}_{II}$ and ${\rm{Im}} \, a^{\bar{\ell}}_{II}$ 
from cuts selecting a lepton or an anti-lepton of flavour $f$ comes only from the diagrams discussed in the previous paragraph 
and can be read off equations \eqref{match1} and \eqref{match2} by simply performing the replacement~\eqref{replacement}. 

As discussed in section~\ref{sec:direct2}, the Majorana neutrino of type $2$, if heavier than the Majorana neutrino 
of type $1$, has an additional source of CP asymmetry whose ultimate origin is the kinematically allowed transition 
$\nu_{R,2} \to \nu_{R,1} + $ Higgs boson. This asymmetry is described in the EFT by the diagrams shown in figure~\ref{fig:DeltaEFT}.
The only difference with the unflavoured case is that we now require for the cut to select a lepton (or anti-lepton) with a specific flavour $f$. 
Hence the cut stands for $-3(F^*_{fI} F_{fJ})\lambda/(8\pi M)$ (or $-3(F^{*}_{fJ} F_{fI})\lambda/(8\pi M)$ in the anti-leptonic case), 
where $I$ is the type of outgoing and $J$ the type of incoming neutrino.
Going through the same derivation as in section~\ref{sec:direct2}, we find
\begin{equation}
\Delta\Gamma_{f2,{\rm direct}}^{\rm mixing} = 
\frac{{\rm{Im}}\left[ (F^{*}_1 F_2) (F^{*}_{f1} F_{f2})\right] + {\rm{Im}}\left[ (F_1 F^{*}_2) (F^{*}_{f1} F_{f2})\right]}{16 \pi^2} \lambda \frac{T^2\Delta}{M^2}.
\label{gammaDelta_flavour}
\end{equation}
The quantity $\Delta\Gamma_{f2,{\rm direct}}^{\rm mixing}$ is the equivalent of $\Delta\Gamma_{2,{\rm direct}}^{\rm mixing}$ in the flavoured case. 
It reduces to $\Delta\Gamma_{2,{\rm direct}}^{\rm mixing}$, given in \eqref{gammaDelta}, when summed over the flavours $f$.

Rewriting the thermal contributions to the direct CP asymmetry given in \eqref{CPnu1} and \eqref{CPnu2} 
for the flavoured case through \eqref{replacement} and adding to the CP asymmetry of the Majorana neutrino of type 2 
the contribution in \eqref{gammaDelta_flavour} proportional to ${\rm{Im}}\left[ (F_1 F^{*}_2) (F^{*}_{f1} F_{f2})\right]$  
gives at order $T^2/M^2$ and at order $\Delta/M$ 
\begin{eqnarray}
\epsilon^{T}_{f1,{\rm direct}} &=& \frac{{\rm{Im}}\left[  (F^{*}_{1}F_{2})(F^{*}_{f1}F_{f2})\right] }{8 \pi |F_{1}|^2}  \left(  \frac{T}{M} \right)^2 
\left\lbrace   \lambda \left[ 2-\ln 2+\left( 1-3\ln 2 \right) \frac{\Delta}{M}\right]  \right.
\nonumber \\
&& 
\left.
- \frac{g^2}{16}\left[ 2- \ln2 +\left( 3 - 5 \ln 2\right) \frac{\Delta}{M}  \right]  
- \frac{g'^2}{48}\left[ 4- \ln2 +\left( 1 - 5 \ln 2\right) \frac{\Delta}{M}  \right]  \right\rbrace,
\label{CPnu1_flavour}
\end{eqnarray}
and 
\begin{eqnarray}
\epsilon^{T}_{f2,{\rm direct}} &=& -\frac{{\rm{Im}}\left[ (F^{*}_{1}F_{2})(F^{*}_{f1}F_{f2}) \right] }{8 \pi |F_{2}|^2}  \left(  \frac{T}{M} \right)^2 
\left\lbrace   \lambda \left[ 2-\ln 2-\left( 9 - 5\ln 2 \right) \frac{\Delta}{M}\right]  \right.
\nonumber \\
&& 
\left.
- \frac{g^2}{16}\left[ 2- \ln2 - 7 \left( 1 - \ln 2\right) \frac{\Delta}{M}  \right]  
- \frac{g'^2}{48}\left[ 4- \ln2 -\left( 9 - 7 \ln 2\right) \frac{\Delta}{M}  \right]  \right\rbrace
\nonumber\\
&& + \frac{{\rm{Im}}\left[ (F_{1}F^{*}_{2})(F^{*}_{f1}F_{f2}) \right] }{2 \pi |F_{2}|^2}  \left(  \frac{T}{M} \right)^2\lambda\frac{\Delta}{M} \,.
\label{CPnu2_flavour}
\end{eqnarray}

Finally, the thermal corrections to the indirect CP asymmetry are easily computed in the EFT. 
The analysis carried out in section~\ref{sec:indirect} is valid also in the flavoured regime. 
The thermal corrections to the indirect CP asymmetry have the same form as \eqref{indirect1T} and \eqref{indirect2T}, 
namely for the two neutrino species
\begin{equation}
\epsilon_{f1,{\rm indirect}}^{T}= -\frac{\epsilon_{f1,{\rm indirect}}^{T=0}}{3} \,\left(|F_2|^2-|F_1|^2\right)\,
\frac{M\Delta}{\Delta^2 + (\Gamma_{22}^{T=0}-\Gamma_{11}^{T=0})^2/4}\,\frac{T^2}{M^2}\,,
\label{indirect1T_fla}
\end{equation}
and
\begin{equation}
\epsilon_{f2,{\rm indirect}}^{T}= -\frac{\epsilon_{f2,{\rm indirect}}^{T=0}}{3} \,\left(|F_2|^2-|F_1|^2\right)\,
\frac{M\Delta}{\Delta^2 + (\Gamma_{22}^{T=0}-\Gamma_{11}^{T=0})^2/4}\,\frac{T^2}{M^2}\,.
\label{indirect2T_fla}
\end{equation}
Note that the first factor in the right-hand side of each asymmetry is the flavoured indirect CP asymmetry at $T=0$
computed in \eqref{indirect1T0_fla} and \eqref{indirect2T0_fla}.

\section{Conclusions}
\label{sec:concl}
In the framework of an extension of the Standard Model that includes two 
generations of heavy Majorana neutrinos with nearly degenerate masses $M$ and $M+\Delta$,  
and coupled only to the SM Higgs boson and lepton doublets via Yukawa interactions, see \eqref{eq3}, 
we have computed the leading thermal corrections to the direct and indirect CP asymmetries 
for neutrino decays into leptons and anti-leptons.
In order to describe a condition that occurred in the early universe, 
we have assumed the SM particles to form a plasma whose temperature $T$ is larger 
than the electroweak scale but smaller than $M$.
Non-vanishing complex phases of the Yukawa couplings originate a CP asymmetry and 
the condition $T\ll M$ puts the Majorana neutrino out of chemical equilibrium.
The main original results of the paper are equations \eqref{CPnu1} and \eqref{CPnu2} for the thermal corrections to the direct CP asymmetry, 
and equations \eqref{indirect1T} and \eqref{indirect2T} for the thermal corrections to the indirect CP asymmetry. 
The corresponding equations for the flavoured case are \eqref{CPnu1_flavour}, \eqref{CPnu2_flavour}
\eqref{indirect1T_fla} and \eqref{indirect2T_fla} respectively.
We have computed the CP asymmetries up to first order in the neutrino mass difference $\Delta \ll M$. 
Moreover, the indirect CP asymmetry has been computed assuming that the mixing can be treated perturbatively. 
Besides this the results are valid in a wide range of parameters. 
In the resonant case ($\Delta$ of the order of the difference of the widths) 
the indirect asymmetry may be the dominant mechanism for the production of a CP asymmetry.

Thermal corrections to the CP asymmetry arise at order $F^4$ in the Yukawa couplings.
Corrections to the direct CP asymmetry are further suppressed by one SM coupling.
Hence the calculation of the thermal effects to the direct CP asymmetry is a three-loop 
calculation in the fundamental theory \eqref{eq3}. We have performed the calculation in 
the effective field theory framework introduced in~\cite{Biondini:2013xua}, which is valid for $T\ll M$. 
The three-loop thermal calculation of the original theory splits into the calculation of the 
imaginary parts of two-loop diagrams that match the Wilson coefficients of the EFT \eqref{eq:efflag}, 
a calculation that can be performed in vacuum, and the calculation of a thermal one-loop diagram in the EFT.
In its range of applicability, the EFT framework provides, therefore, a significantly simpler method of calculation.
The same formalism may prove to be a useful tool to calculate the CP asymmetry 
also in other arrangements of the heavy-neutrino masses, such as a hierarchically ordered 
neutrino mass spectrum, where direct and indirect CP asymmetries are of comparable size.
The EFT \eqref{eq:efflag} is also the natural starting point for establishing the rate equations 
for the time evolution of the particle densities in the regime where the Majorana neutrinos are non-relativistic.
A first study of the non-relativistic approximation for the rate equations can be found in~\cite{Bodeker:2013qaa}.

There are some critical issues about the results presented here that should be mentioned and be possibly the subject of further investigations. 
The results rely on a strict expansion in $T/M$. The range of applicability of this expansion has been investigated in~\cite{Laine:2013lka} 
for the neutrino production rate by comparing with exact results.
Although the expansion converges well, its agreement with the exact result appears to happen at relatively small temperatures.
A similar behaviour could be also for the CP asymmetry.
We investigate this issue and provide a computational scheme that may solve it in appendix~\ref{AppendixD}.

Another question is how the corrections in $T/M$ compare with the yet unknown radiative corrections to the CP asymmetry at zero temperature. 
First, we note that for the indirect CP asymmetry, which is the dominant part of the asymmetry in particular 
for the resonant case or close to it, the computed $(T/M)^2$ corrections are not suppressed by the SM couplings. 
Hence they are likely to be larger than or of the same size as radiative corrections for a wide range of temperatures.
Second, we observe that thermal corrections to the direct CP asymmetry, which are suppressed in the SM couplings,  
are indeed of relative size $\lambda (T/M)^2$ and $(3g^2+g'^2)(T/M)^2$ 
(cf.~with \eqref{CPnu1} and \eqref{CPnu2} or \eqref{CPnu1_flavour} and \eqref{CPnu2_flavour}). 
These should be compared with radiative corrections of possible relative size $\lambda/\pi^2$, $|\lambda_t|^2/\pi^2$ or $(3g^2+g'^2)/\pi^2$
(cf.~with the radiative corrections to the production rate in~\cite{Salvio:2011sf}).
The factor $1/\pi^2$ is typical of radiative corrections, but absent in thermal corrections.
The two are of comparable size for $T/M \sim 1/\pi$, which is inside the range of convergence of the expansion in $T/M$.
Clearly radiative corrections are a missing ingredient for a complete quantitative evaluation of the CP asymmetry.
Following the above discussion, their evaluation seems most needed when the CP asymmetry 
is dominated by direct contributions and at lower temperatures.

At relative order $(T/M)^2$ only the Higgs self-coupling, $\lambda$, 
and the SU(2)$_L\times$U(1)$_Y$ gauge couplings, $g$ and $g'$, enter the expression of the CP asymmetry. 
Higher-order operators in the $1/M$ expansion have not been considered in this work.  
However, higher-order operators, most importantly the dimension seven operators described in~\cite{Biondini:2013xua}, 
may contribute to the CP asymmetry as well. The power counting of the EFT
shows that they can induce thermal corrections that scale like $g_{\hbox{\tiny SM}}(T/M)^4$, 
where $g_{\hbox{\tiny SM}}$ is understood as either $\lambda$, $(3g^2+g'^2)$ or the top Yukawa coupling $|\lambda_t|^2$. 
Even though these corrections are further suppressed in the expansion in $T/M$, 
the particular values of the SM couplings at high energies can make 
$g_{\hbox{\tiny SM}}(T/M)^4$ corrections numerically comparable with or larger 
than those calculated at order $(T/M)^2$  and presented in this work. 
As a reference, at a scale of $10^4$~TeV the Higgs self coupling is about $\lambda\approx 0.02$, 
the top Yukawa coupling is about $|\lambda_t|^2\approx 0.4$ and $(3g^2+g'^2) \approx 1.2$,
whereas  at a scale of $1$~TeV $\lambda\approx 0.1$, $|\lambda_t|^2\approx 0.7$ and $(3g^2+g'^2) \approx 1.6$~\cite{Rose15,Buttazzo:2013uya}. 
To shape better this issue the effect of, at least, some higher-order operators should be calculated.

\acknowledgments

We thank Marco Drewes, Bj\"orn Garbrecht, Alexander Kartavtsev, Emiliano Molinaro, Enrico Nardi and Luigi delle Rose for several discussions, 
and Vladyslav Shtabovenko for checking some of the integrals.
We thank the Mainz Institute for Theoretical Physics for giving us the opportunity 
to organize the institute \emph{Jet Particles and transport properties in collider and
cosmological environments} in summer 2014 during which some of this work was presented.
We acknowledge financial support from the DFG cluster of excellence
\emph{Origin and structure of the universe} (www.universe-cluster.de). 
M.A.E. acknowledges support from the European Research Council under the Advanced Investigator Grant ERC-AD-267258.

\appendix

\section{Majorana neutrino propagators}
\label{AppendixA}
In this section we review the expressions for the relativistic propagators of a Majorana fermion 
and the corresponding non-relativistic version~\cite{Biondini:2013xua}. 
If $\psi_I$ is a spinor describing a relativistic Majorana particle, then
\begin{equation}
\psi_I=\psi_I^{c}=C  \bar{\psi}_I^{\,T} \, ,
\label{eq:Majodef}
\end{equation}
where $\psi_I^{c}$ denotes the charge-conjugate spinor and $C$ the charge-conjugation matrix that satisfies $C^\dagger=C^T=C^{-1}=-C$ 
and $C\,\gamma^{\mu\,T}\,C = \gamma^\mu$. The relativistic propagators for a free Majorana particle are:
\begin{eqnarray}
\langle 0 | T( \psi_I^{\alpha} (x) \bar{\psi}_I^{\beta} (y)  )| 0 \rangle &=& 
i \int \frac{d^{4}p}{(2 \pi)^{4}} \, \frac{(\slashed{p}+M_I)^{\alpha \beta}}{p^{2}-M_I^{2}+i\epsilon}  \, e^{-ip \cdot (x-y)} \,,
\label{A1}
\\
\langle 0 | T( \psi_I^{\alpha}(x) \psi_I^{\beta} (y) )| 0 \rangle &=& 
-i \int \frac{d^{4}p}{(2 \pi)^{4}} \, \frac{\left[  (\slashed{p}+M_I) C \right]^{\alpha \beta}}{p^{2}-M_I^{2}+i\epsilon} \,  e^{-ip \cdot (x-y)} \,,
\label{A2}
\\
\langle 0 | T( \bar{\psi}_I^{\alpha} (x) \bar{\psi}_I^{\beta} (y)  ) | 0 \rangle &=& 
-i \int \frac{d^{4}p}{(2 \pi)^{4}} \, \frac{ \left[  C (\slashed{p}+M_I) \right]^{\alpha \beta}}{p^{2}-M_I^{2}+i\epsilon} \, e^{-ip \cdot (x-y)} \,,
\label{A3}
\end{eqnarray}
where $\alpha$ and $\beta$ are Lorentz indices and $T$ stands for the
time-ordered product. The expression for the non-relativistic Majorana
propagator in the EFT (\ref{eq:efflag}) can be obtained by projecting (\ref{A1})-(\ref{A3}) on the small components of the Majorana fields. 
Putting $p^{\mu}=Mv^{\mu}+k^{\mu}$, where $k^2 \ll M^2$, we obtain in the large $M$ limit
\begin{equation}
\langle 0 | T ( N_1^{\alpha}(x) \bar{N}_1^{\beta}(y) )  | 0 \rangle 
= \left( \frac{1+\slashed{v}}{2}\right)^{\alpha \beta} \int \frac{d^{4}k}{(2 \pi)^{4}} \, e^{-ik\cdot(x-y)}\, \frac{i}{v\cdot k +i\epsilon}  \,  ,
\label{effpropagator}
\end{equation} 
and 
\begin{equation}
\langle 0 | T ( N_2^{\alpha}(x) \bar{N}_2^{\beta}(y) )  | 0 \rangle 
= \left( \frac{1+\slashed{v}}{2}\right)^{\alpha \beta} \int \frac{d^{4}k}{(2 \pi)^{4}} \, e^{-ik\cdot(x-y)}\, \frac{i}{v\cdot k -\Delta +i\epsilon}  \,  ,
\label{effpropagator2}
\end{equation} 
where $M_1=M$ and $\Delta = M_2-M_1$. The other possible time-ordered combinations vanish.

\section{Matching the asymmetry}
\label{AppendixB}
In this appendix, we compute the matching coefficients in \eqref{match1} and \eqref{match2}. 
They are obtained by matching matrix elements calculated in the fundamental theory with matrix elements in the EFT. 
The fundamental theory is \eqref{eq3}. It contains the SM with unbroken gauge group SU(2)$_L\times$U(1)$_Y$, whose Lagrangian reads
\begin{eqnarray}
\mathcal{L}_{\hbox{\tiny {SM}}} &=&
\bar{L}_{f} P_R\, i \slashed{D} \, L_{f} + \bar{Q}P_R\, i\slashed{D} \, Q + \bar{t}P_L\, i\slashed{D} \, t -\frac{1}{4}W_{\mu\nu}^aW^{a\,\mu\nu}  -\frac{1}{4}F_{\mu\nu}F^{\mu\nu}
\nonumber\\
&& + \left( D_{\mu} \phi \right)^{\dagger}\left( D^{\mu} \phi \right)  - \lambda \left( \phi^{\dagger}\phi \right)^2 
- \lambda_{t}    \, \bar{Q} \, \tilde{\phi} \, P_{R} t
- \lambda^{*}_{t} \, \bar{t} P_{L} \, \tilde{\phi}^{\dagger} \, Q  + \dots \,.
\label{SMlag}
\end{eqnarray}
The dots stand for terms that can be neglected in our calculation, e.g., terms with right-handed leptons or light quarks. 
The covariant derivative is given by 
\begin{equation}
D_{\mu}=\partial_{\mu} -igA^{a}_{\mu} \tau^{a} -ig'YB_{\mu} \, ,
\label{SMCov}
\end{equation}
where $\tau^{a}$ are the SU(2)$_L$ generators and $Y$ is the hypercharge
($Y=1/2$ for the Higgs boson, $Y=-1/2$ for left-handed leptons). 
The fields $L_{f}$ are the SU(2)$_L$ lepton doublets with flavour $f$, $Q^T=(t,b)$ is
the heavy-quark SU(2)$_L$ doublet, $A^{a}_{\mu}$ are the SU(2)$_L$ gauge
fields, $B_{\mu}$ the U(1)$_Y$ gauge fields and $W^{a\,\mu\nu}$,
$F_{\mu\nu}$ the corresponding field strength tensors, $\phi$ is the
Higgs doublet and $t$ is the top quark field.  The couplings $g$, $g'$,
$\lambda$ and $\lambda_t$ are the SU(2)$_L$ and U(1)$_Y$ gauge couplings, the
four-Higgs coupling and the top Yukawa coupling respectively. 
Because in the matching we integrate out only high-energy modes, we can set to zero any low-energy scale appearing in loops.  
Especially, as discussed in the main body of the paper, we can set to zero the temperature.  
As a consequence, loop diagrams on the EFT side of the matching vanish in dimensional regularization because they are scaleless. 
Dimensional regularization is used for loop calculations throughout the paper.
The operators in the EFT~\eqref{eq:efflag} that we need to match are 
\begin{equation}
\frac{a_{IJ}}{M_I}\bar{N}_IN_J \phi^{\dagger}\phi.
\label{aijoperator}
\end{equation}
Hence we need to consider four-field matrix elements involving two Majorana and two Higgs fields. 
The effective interaction with either leptons, quarks or gauge bosons in the plasma is described 
by operators that are further suppressed in the $1/M$ expansion. We do not consider such operators
in this work since we calculate corrections to the CP asymmetry of relative order $T^2/M^2$, 
whereas the neglected ones induce corrections that are at least of order~$T^4/M^4$.

We perform the matching in the reference frame $v^{\mu}=(1,\vec{0}\,)$, 
where we assume both the Majorana neutrino and the plasma to be at rest.  
Since we are interested in the imaginary parts of the Wilson coefficients, 
we evaluate the imaginary parts of $-i {\mathcal{D}}$, where ${\mathcal{D}}$ are generic Feynman diagrams amputated of the external legs.  
Moreover we may choose the incoming and outgoing SM particles to have vanishing momentum, because their momentum is
assumed to be much smaller than $M$, and we do not match onto derivative operators. 
(An exception are diagrams with pinch singularities where we set the momentum to zero after the cancellation of the singularities).

\subsection{Cutting rules}
\label{appcutting}
A way of computing the imaginary part of $-i {\mathcal{D}}$, where ${\mathcal{D}}$ is a Feynman diagram, is by means of cutting rules.
Here we describe briefly the cutting rules at zero temperature and the notation that we will use; we also illustrate them with an example. 
We refer to~\cite{Cutkosky:1960sp,Remiddi:1981hn,Bellac} for some classical presentations and to~\cite{Denner:2014zga} for a more recent one 
suited to include complex masses and couplings.

\begin{figure}[ht]
\centering
\includegraphics[scale=0.55]{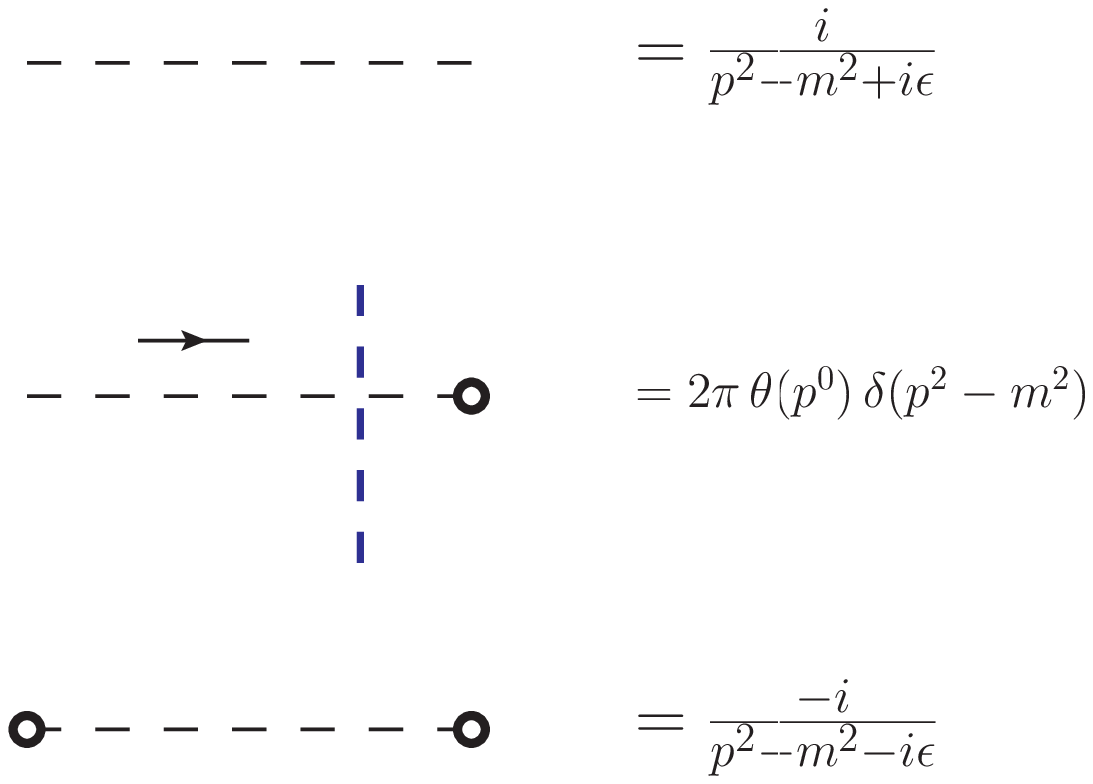}
\caption{The relevant cutting rules for a scalar propagator at zero temperature 
in the convention of~\cite{Bellac}. The momentum direction is represented by the arrow. 
The blue thick dashed line stands for the cut. Vertices on the right of the cut are circled. 
Circled vertices have opposite sign than non-circled vertices.}
\label{fig:cutting} 
\end{figure}

At the core of the method is the cutting equation, which relates ${\rm{Im}}(-i {\mathcal{D}})$ with cut diagrams of ${\mathcal{D}}$.
It reads 
\begin{equation}
{\rm{Im}}(-i \mathcal{D})= - {\rm{Re}}(\mathcal{D})=\frac{1}{2} \sum_{\rm cuts} \mathcal{D} \, .
\label{cuttingEquation}
\end{equation} 
A cut diagram consists in separating the Feynman diagram into two disconnected diagrams by putting on shell some of its internal propagators. 
The cut is typically represented by a line ``cutting'' through these propagators: in our case it is a blue thick dashed line.
Vertices on the right of the cut are circled.  Circled vertices have opposite sign than uncircled vertices. 
We can have three types of propagators. Propagators between two circled vertices, propagators between uncircled vertices 
and propagators between one circled and one uncircled vertex. This last situation occurs when the cut goes through the propagator.
The expressions for these three propagators are shown in the case of a scalar particle in figure~\ref{fig:cutting};
the extension to fermions and gauge bosons is straightforward.
Note that when the cut goes through the propagator the particle is put on shell.
The sum in \eqref{cuttingEquation} extends over all possible cuts of the diagram ${\mathcal{D}}$. 

As an example, we show how to obtain the imaginary part of the Wilson coefficient of the operator~\eqref{aijoperator} 
in the case of just one neutrino generation. We call this single Wilson coefficient $a$.
It was first derived in~\cite{Biondini:2013xua} without using cutting rules.
Cutting rules have the advantage that they allow to disentangle the contribution coming from the decay into a lepton,
which we call ${\rm Im}\,a^\ell$, from the contribution coming from the decay into an anti-lepton, which we call ${\rm Im}\,a^{\bar{\ell}}$.
The coefficient ${\rm Im}\,a$ is at leading order the sum of these two contributions: ${\rm Im}\,a= {\rm Im}\,a^\ell + {\rm Im}\,a^{\bar{\ell}}$.
It can be obtained by matching the following matrix element of the fundamental theory
\begin{equation}
-i \left.\int d^{4}x\,e^{i p \cdot x} \int d^{4}y \int d^{4}z\,e^{i q \cdot (y-z)}\, 
\langle \Omega | T(\psi^{\mu}(x) \bar{\psi}^{\nu }(0) \phi_{m}(y) \phi_{n}^{\dagger}(z) )| \Omega \rangle
\right|_{p^\alpha =(M + i\epsilon,\vec{0}\,)},
\label{matrix}
\end{equation} 
with the corresponding matrix element of the EFT. The field $\psi$ identifies the only Majorana 
neutrino field available in this case, $\mu$ and $\nu$ are Lorentz indices and $m$ and $n$ SU(2)$_L$ indices.

\begin{figure}[ht]
\centering
\includegraphics[scale=0.5]{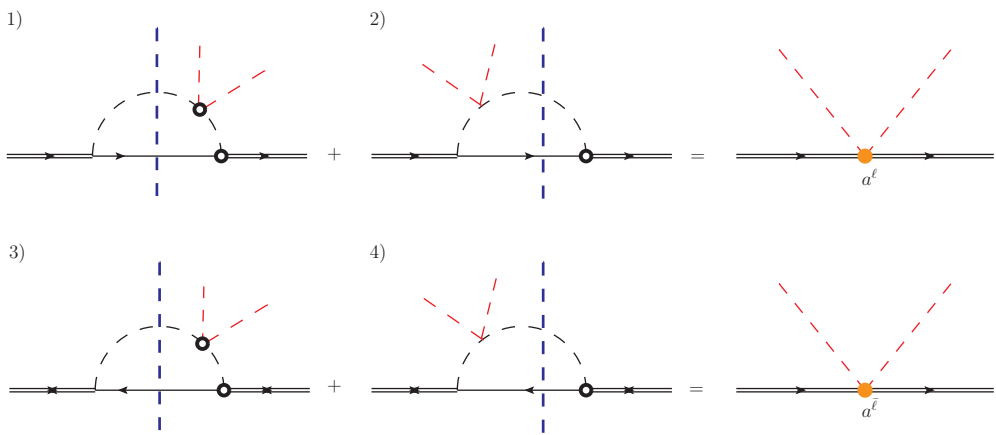}
\caption{Diagrams in the full theory contributing to the Majorana neutrino-Higgs boson dimension-five operator. 
On the left-hand side are the diagrams in the full theory, whereas on the right-hand side are the diagrams in the EFT.
As in figure~\ref{fig:treeMatch} and in the rest of the paper, red dashed lines indicate external Higgs bosons  
with a soft momentum much smaller than the mass of the Majorana neutrino. The cuts on the diagrams 
in the fundamental theory are explicitly shown.}
\label{Fig0} 
\end{figure}

When computing matrix elements involving Majorana fermions, one has to consider 
that the relativistic Majorana field may be contracted in more ways than if it was a Dirac field, 
this reflecting the indistinguishability of the Majorana particle and anti-particle.
The different contractions give rise to the different propagators listed in \eqref{A1}-\eqref{A3}.
When contracting the Majorana fields in \eqref{matrix} according to \eqref{A1}, one obtains at leading order 
\begin{eqnarray}
\left[\hat{P}\left( -i \mathcal{D}\right)\hat{P}\right]^{\mu\nu}
= 6 |F|^2 \lambda \, \delta_{mn} \,  \int \frac{d^{4} \ell }{(2\pi)^{4}}  
\left( \hat{P} \, P_L \slashed{\ell} \, \hat{P} \right)^{\mu \nu} \frac{i}{\ell^{2}+i \epsilon} 
\left( \frac{i}{(M v - \ell )^2+i \epsilon} \right)^{2} ,
\label{3a}
\end{eqnarray}
where we have dropped all external propagators and $\mathcal{D}$ is the amputated (uncut) diagram shown in the upper raw and left-hand side of figure~\ref{Fig0}.
The external heavy-neutrino propagators reduce in the non-relativistic limit and in the rest frame 
to a matrix proportional to $\hat{P}=(1+\gamma^{0})/2$ (see \eqref{effpropagator}).
We have kept the matrix $\hat{P}$ on the left- and right-hand side of \eqref{3a}, because it helps projecting 
out the contributions relevant in the heavy-neutrino mass limit, e.g., $ \hat{P} \, P_L \, \hat{P} = \hat{P}/2$.
After projection, also the matrix $\hat{P}$ may be eventually dropped from the left- and right-hand side of the matching equation.
The internal loop momentum is $\ell^{\mu}$, $Mv^{\mu}=(M,\vec{0})$ is the neutrino momentum in the rest frame and $|F|^2=\sum_{f} F^{*}_{f} F_{f}$. 

The diagram $\mathcal{D}$ admits two cuts labeled $1)$ and $2)$ and shown in the upper raw and left-hand side of figure~\ref{Fig0}.
Both cuts select a final state made of a lepton and, therefore, contribute to~$a^\ell$.
Using \eqref{cuttingEquation} and the cutting rules we obtain for the two cuts:
\begin{eqnarray}
\left[\hat{P}\, {\rm{Im}}(-i\mathcal{D}^{\ell}_{1,\hbox{\tiny fig.\ref{Fig0}}})\hat{P}\right]^{\mu\nu}
=3 |F|^2 \lambda \,(-1)^2 \, \delta_{mn} \, \int \frac{d^{4} \ell}{(2\pi)^{4}}    
\left( \hat{P} \, P_L \slashed{\ell} \, \hat{P} \right)^{\mu \nu}\,2\pi \theta(\ell^{0}) \delta(\ell^2) 
\nonumber \\ 
\times 2\pi\theta(M-\ell^{0})\delta((Mv-\ell)^2) \frac{-i}{(M v - \ell)^2  -i \epsilon} ,
\label{cut1}
\end{eqnarray}
\begin{eqnarray}
\left[\hat{P}\, {\rm{Im}}(-i\mathcal{D}^{\ell}_{2,\hbox{\tiny fig.\ref{Fig0}}})\hat{P}\right]^{\mu\nu}
=3 |F|^2 \lambda \, (-1)  \, \delta_{mn} \,  
\int \frac{d^{4}\ell}{(2\pi)^{4}} \left( \hat{P} \, P_L \slashed{\ell} \, \hat{P} \right)^{\mu \nu}\,2\pi  \theta(\ell^{0}) \delta(\ell^2)  
\nonumber \\
\times 2\pi\theta(M-\ell^{0})\delta((Mv-\ell)^2) \frac{i}{(M v - \ell )^2  +i \epsilon} .
\label{cut2}
\end{eqnarray}
Both ${\rm{Im}}(-i\mathcal{D}^{\ell}_{1,\hbox{\tiny fig.\ref{Fig0}}})$ and ${\rm{Im}}(-i\mathcal{D}^{\ell}_{2,\hbox{\tiny fig.\ref{Fig0}}})$ have a pinch singularity 
whose origin is the soft limit of the Higgs momentum pair. A way to regularize the singularity 
is to give a small finite momentum to the Higgs pair and set it to zero after cancellation of the singularity. 
The singularity cancels in the sum of the two cuts, which reads
\begin{equation}
{\rm{Im}}(-i\mathcal{D}^{\ell}_{1,\hbox{\tiny fig.\ref{Fig0}}}) + {\rm{Im}}(-i\mathcal{D}^{\ell}_{2,\hbox{\tiny fig.\ref{Fig0}}})=
-\frac{3}{16 \pi M} |F|^2 \lambda \, \delta^{\mu \nu} \delta_{mn}, 
\label{E}
\end{equation} 
where we have used for the amputated Green function the same indices used for the unamputated one, a convention that we will keep in the following. 

When contracting the Majorana fields in~\eqref{matrix} according to~\eqref{A2} and~\eqref{A3} 
one obtains at leading order a contribution encoded in the diagram shown in the lower raw and left-hand side of figure~\ref{Fig0}.
The expression for this diagram is the same as the one in~\eqref{3a} up to an irrelevant change $P_L \to P_R$
(the expression is also unsensitive to the change $F_f  \leftrightarrow F_f^{*}$).
The diagram admits two cuts labeled $3)$ and $4)$ and shown in the lower raw and left-hand side of figure~\ref{Fig0}.
Both cuts select a final state made of an anti-lepton and, therefore, contribute to  $a^{\bar{\ell}}$.
The contributions from these two cuts are the same as the ones in \eqref{cut1} and \eqref{cut2} and give eventually the same result for the sum
\begin{equation}
{\rm{Im}}(-i\mathcal{D}^{\bar{\ell}}_{3,\hbox{\tiny fig.\ref{Fig0}}}) + {\rm{Im}}(-i\mathcal{D}^{\bar{\ell}}_{4,\hbox{\tiny fig.\ref{Fig0}}})=
-\frac{3}{16 \pi M} |F|^2 \lambda \, \delta^{\mu \nu} \delta_{mn}.
\label{Ebis}
\end{equation} 

Comparing \eqref{E} and \eqref{Ebis} with the corresponding expressions in the EFT, which are 
$({\rm{Im}} \,a^{\ell}/M) \, \delta^{\mu \nu} \delta_{mn}$ and $({\rm{Im}} \,a^{\bar{\ell}}/M) \, \delta^{\mu \nu} \delta_{mn}$ respectively, one obtains
\begin{eqnarray}
&& {\rm{Im}} \, a^{\ell} = {\rm{Im}} \, a^{\bar{\ell}} = -\frac{3}{16\pi}|F|^{2}\lambda ,
\label{alalbar}\\
&& {\rm{Im}} \, a={\rm{Im}} \, a^{\ell}+{\rm{Im}} \, a^{\bar{\ell}}= -\frac{3}{8\pi}|F|^{2}\lambda .
\label{atot}
\end{eqnarray} 
Equation \eqref{atot} agrees with the result found in~\cite{Biondini:2013xua}.

\subsection{Matching diagrams with four-Higgs interaction}
\label{appHiggs}
In order to derive \eqref{match1}, we compute in the fundamental theory the matrix element 
\begin{equation}
-i \left.\int d^{4}x\,e^{i p \cdot x} \int d^{4}y \int d^{4}z\,e^{i q \cdot (y-z)}\, 
\langle \Omega | T(\psi^{\mu}_{1}(x) \bar{\psi}^{\nu }_{1}(0) \phi_{m}(y) \phi_{n}^{\dagger}(z) )| \Omega \rangle
\right|_{p^\alpha =(M + i\epsilon,\vec{0}\,)}.
\label{B1}
\end{equation} 
The matrix element is similar to \eqref{matrix}, but now in a theory with two types of heavy Majorana neutrinos.
External neutrinos are of type 1, whereas neutrinos of type 2 appear only as intermediate states. 
The result can be extended straightforwardly to the case of external neutrinos of type 2, leading to \eqref{match2}. 
The matrix element describes a $2 \rightarrow 2$ scattering between a heavy Majorana neutrino 
of type 1 at rest and a Higgs boson carrying momentum $q^\mu$. 
Since the momentum $q^\mu$ is much smaller than the neutrino mass 
and we are not matching derivative operators, $q^\mu$ can be set to zero in the matching.
Here, we compute the diagrams contributing to \eqref{B1} that enter the matching of $a^{\ell}_{11}$ 
(and $a^{\bar{\ell}}_{11}$) up to first order in $\lambda$ and are relevant for the direct CP asymmetry;
in the next section, we will compute the diagrams of order $g^2$ and $g'^2$.
It may be useful to cast the diagrams into three different typologies as we will do in the following.
All diagrams are understood as amputated of their external legs.

\begin{figure}[ht]
\centering
\includegraphics[scale=0.55]{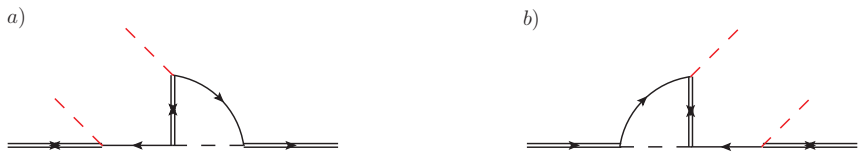}
\caption{Diagrams contributing to $a^\ell_{II}$ at order $F^4$. One diagram is the complex conjugate of the other.}
\label{fig:F4}
\end{figure}

A first class of diagrams is obtained by opening-up a Higgs line in the two-loop diagrams of figure~\ref{Fig2}. 
These diagrams are of order $F^4$. The subset contributing to $a^\ell_{II}$ is shown in figure~\ref{fig:F4}.
Diagrams $a)$ and $b)$ are one the complex conjugate of the other; their sum is real.
By cutting the loops so to bring one lepton on shell and summing both diagrams the result is proportional 
to the Yukawa coupling combination ${\rm{Re}}\left[ (F_1^{*}F_J)^2\right]$ only.
The reason is that, after the cuts, the diagrams do not contain loops anymore and cannot develop any additional complex phase. 
If we consider the subset of diagrams contributing to $a^{\bar{\ell}}_{II}$, which are diagrams where the anti-lepton can be put on shell, 
we obtain through a similar argument that the sum of diagrams is proportional again to the Yukawa coupling combination ${\rm{Re}}\left[(F_1^{*}F_J)^2\right]$. 
It follows that the matching coefficients obtained for leptons and anti-leptons and the corresponding leptonic and anti-leptonic widths 
cancel in the difference. One-loop diagrams of order $F^4$ with two external Higgs bosons do not contribute to the direct CP asymmetry.  

\begin{figure}[ht]
\centering
\includegraphics[scale=0.5]{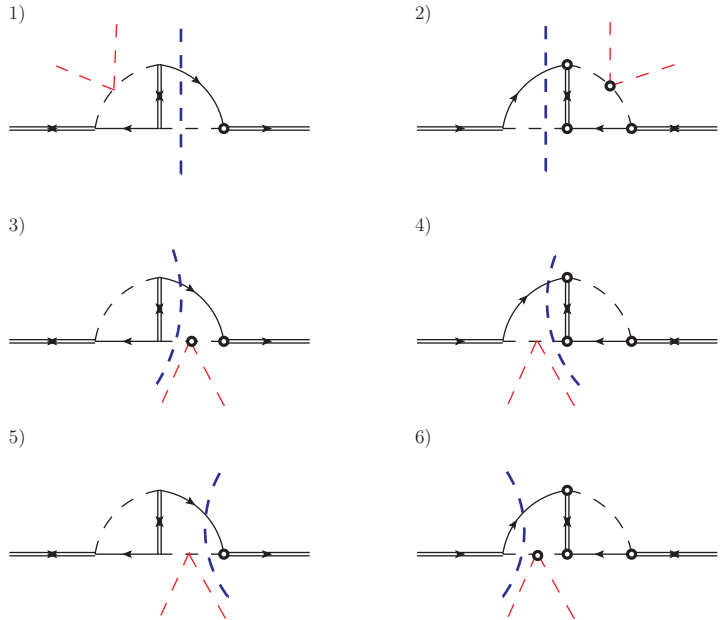}
\caption{Diagrams contributing to $a^\ell_{II}$ and $a^{\bar{\ell}}_{II}$ at order $F^4\lambda$.
The cuts through leptons are explicitly shown and implemented according to the rules of figure~\ref{fig:cutting}.}
\label{appendix0} 
\end{figure}

A second class of diagrams is obtained by attaching a four-Higgs vertex to an existing Higgs line in the two-loop diagrams of figure~\ref{Fig2}.
These diagrams are of order $F^4\lambda$ and are shown with the relevant cuts in figure~\ref{appendix0}. 
In each raw we show a diagram and its complex conjugate and we draw explicitly the cuts that put a lepton on shell.
This amounts at selecting in all the diagrams in figure~\ref{appendix0}
the decay of a heavy Majorana neutrino into a lepton. 
The decay width into an anti-lepton can be computed by cutting anti-lepton lines. 
In general, the sum of each couple of diagrams in figure~\ref{appendix0}
is a linear combination of the real and the imaginary parts of $(F_{1}^{*}F_{J})^2$. 
The appearance of a term proportional to ${\rm{Im}}\left[(F_1^{*}F_2)^2\right]$
in addition to ${\rm{Re}}\left[(F_1^{*}F_J)^2\right]$ reflects the fact that after the cut we are 
left with a loop that also develops an imaginary part.
For each couple of diagrams, contributions coming from the lepton and the anti-lepton cuts give 
the same terms proportional to  ${\rm{Re}}\left[(F_1^{*}F_J)^2\right]$  but terms 
proportional to  ${\rm{Im}}\left[(F_{1}^{*}F_{2})^2\right]$ with opposite signs, 
since ${\rm{Re}}\left[(F_1^{*}F_J)^2\right] = {\rm{Re}}\left[(F_1F_J^{*})^2\right]$
while ${\rm{Im}}\left[(F_{1}^{*}F_{2})^2\right] = - {\rm{Im}}\left[(F_{1}F_{2}^{*})^2\right]$.
So that, when calculating the CP asymmetry, terms proportional to ${\rm{Re}}\left[(F_1^{*}F_J)^2\right]$ cancel, 
and only those proportional to ${\rm{Im}}\left[(F_{1}^{*}F_{2})^2\right]$ remain.
Hence for each diagram we only need to calculate the terms proportional to ${\rm{Im}}\left[(F_{1}^{*}F_{2})^2\right]$, 
consistently with the discussion in section~\ref{sec:zeroT}.
Up to relative order $\Delta/M$ they are:
\begin{eqnarray}
&&\!\!\!\!\! \!\!\!\!\! 
{\rm{Im}}\, (-i \mathcal{D}^{\ell}_{1,\hbox{\tiny fig.\ref{appendix0}}}) + {\rm{Im}}\, (-i \mathcal{D}^{\ell}_{2,\hbox{\tiny fig.\ref{appendix0}}}) = 
\nonumber \\
&&\hspace{3.7cm}  
\frac{3\,{\rm{Im}}\left[ (F_{1}^{*}F_{2})^2\right]}{(16 \pi)^2M} \lambda \left[ \ln 2 -(1-\ln 2) \frac{\Delta}{M}\right] 
\delta^{\mu \nu} \delta_{mn} + \dots \, ,  
\label{Higgs1} 
\\
&&\!\!\!\!\! \!\!\!\!\! 
{\rm{Im}}\,(-i \mathcal{D}^{\ell}_{3,\hbox{\tiny fig.\ref{appendix0}}}) + {\rm{Im}}\,(-i \mathcal{D}^{\ell}_{4,\hbox{\tiny fig.\ref{appendix0}}}) 
+ {\rm{Im}}\,(-i \mathcal{D}^{\ell}_{5,\hbox{\tiny fig.\ref{appendix0}}}) + {\rm{Im}}\,(-i \mathcal{D}^{\ell}_{6,\hbox{\tiny fig.\ref{appendix0}}}) = 
\nonumber \\
&&\hspace{3.7cm}  
\frac{ 3\,{\rm{Im}}\left[ (F_{1}^{*}F_{2})^2\right] }{(16 \pi)^2M} \lambda\left[ \ln 2 -(1-\ln 2) \frac{\Delta}{M}\right]
\delta^{\mu \nu} \delta_{mn} + \dots \, .
\label{Higgs2}
\end{eqnarray}   
The dots stand for terms proportional to the Yukawa coupling combination ${\rm{Re}}\left[(F_{1}^{*}F_{J})^2\right]$ 
and higher-order terms in the expansion in $\Delta/M$.  
The superscript $\ell$ reminds that we have cut through leptons only; 
as we argued above, the contribution of anti-leptons has opposite sign.
We give the result in \eqref{Higgs2} as the sum of four diagrams to cancel 
a pinch singularity that arises in the soft momentum limit of the Higgs boson. 
This is analogous to the calculation carried out in section~\ref{appcutting}.

\begin{figure}[ht]
\centering
\includegraphics[scale=0.5]{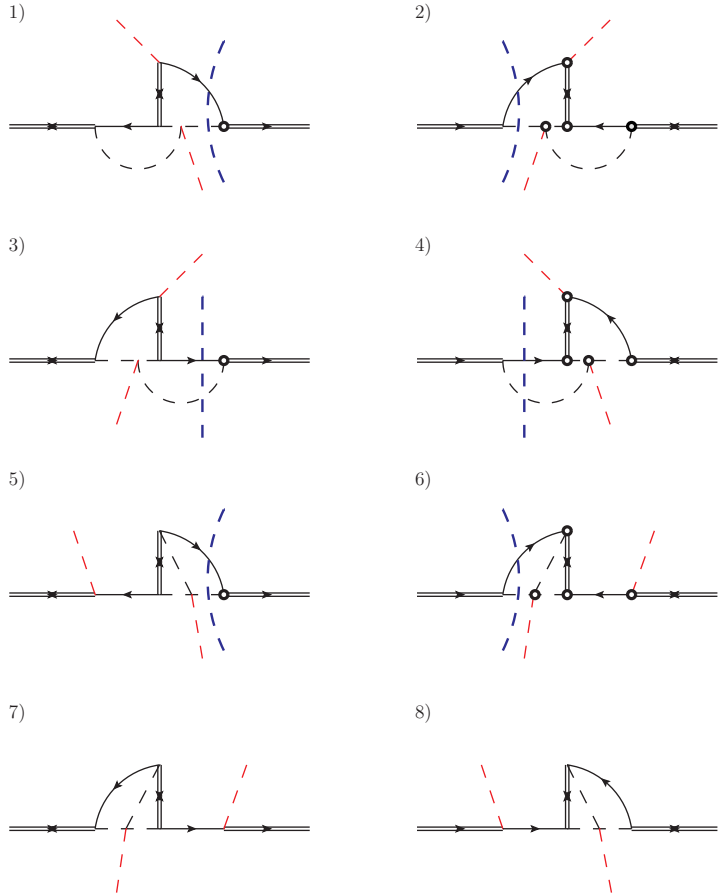}
\caption{Diagrams contributing to $a^\ell_{II}$ and $a^{\bar{\ell}}_{II}$ at order $F^4\lambda$. The cuts through leptons are explicitly shown.}
\label{fig:new_Higgs_direct} 
\end{figure}

Once the four-Higgs vertices are removed, the diagrams of figure~\ref{appendix0} preserve the topology of the $T=0$ two-loop diagrams of figure~\ref{Fig2}.
There is, finally, a third class of diagrams where this topology is not preserved. 
A way to construct them is from the diagrams of figure~\ref{fig:F4} (and the corresponding ones with an anti-lepton in the loop)
by adding a four-Higgs vertex to the internal Higgs line; we show the diagrams with the relevant cuts in figure~\ref{fig:new_Higgs_direct}. 
The results for the cuts through leptons read
\begin{eqnarray}
&&\!\!\!\!\! \!\!\!\!\! 
{\rm{Im}}\, (-i \mathcal{D}^{\ell}_{1,\hbox{\tiny fig.\ref{fig:new_Higgs_direct}}}) + {\rm{Im}}\, (-i \mathcal{D}^{\ell}_{2,\hbox{\tiny fig.\ref{fig:new_Higgs_direct}}}) = 
\nonumber \\
&&\hspace{3.7cm}  
\frac{3\,{\rm{Im}}\left[ (F_{1}^{*}F_{2})^2\right]}{(16 \pi)^2M} \lambda  \left(1-\frac{\Delta}{M} \right)
\delta^{\mu \nu} \delta_{mn} + \dots \, ,  
\label{Higgs3} 
\\
&&\!\!\!\!\! \!\!\!\!\! 
{\rm{Im}}\,(-i \mathcal{D}^{\ell}_{3,\hbox{\tiny fig.\ref{fig:new_Higgs_direct}}}) + {\rm{Im}}\,(-i \mathcal{D}^{\ell}_{4,\hbox{\tiny fig.\ref{fig:new_Higgs_direct}}}) =
\nonumber \\
&&\hspace{3.7cm}  
\frac{3\,{\rm{Im}}\left[ (F_{1}^{*}F_{2})^2\right]}{(16 \pi)^2M} \lambda  \left(1-\frac{\Delta}{M} \right)
\delta^{\mu \nu} \delta_{mn} + \dots \, ,  
\label{Higgs4}
\\
&&\!\!\!\!\! \!\!\!\!\! 
{\rm{Im}}\,(-i \mathcal{D}^{\ell}_{5,\hbox{\tiny fig.\ref{fig:new_Higgs_direct}}}) + {\rm{Im}}\,(-i \mathcal{D}^{\ell}_{6,\hbox{\tiny fig.\ref{fig:new_Higgs_direct}}}) =0\,.
\label{Higgs5}
\end{eqnarray}    

Some remarks, which will be of use also in the following to simplify the calculation, are in order.
First, in the Feynman diagrams, integrals over momentum regions where the intermediate neutrino is on shell do no contribute to the matching. 
Such momentum regions are either kinematically forbidden, if the intermediate neutrino is heavier 
than the initial one, or they are reproduced in the EFT, if the intermediate neutrino is lighter than the initial one
(see diagrams in figure~\ref{fig:DeltaEFT} and the related discussion in section~\ref{sec:direct2}).
In the last case, the momentum is necessarily of order $\Delta$.
Modes with energy or momentum of order $\Delta \ll M$ are still dynamical in the effective theory 
and should not be integrated out with the mass scale (if they are, then they would need to be subtracted 
by computing suitable loops in the effective theory). Second, also momentum regions where three massless particles happen 
to be on-shell and enter the same vertex do not contribute to the matching, because the available phase space vanishes in dimensional regularization.
These general remarks apply in the present case to the diagrams 5) and 6) of figure~\ref{fig:new_Higgs_direct}.
After the cuts through the lepton propagators shown in the diagrams have been implemented, the remaining one-loop diagrams may 
develop an imaginary part only if two of the particles in the loop can be put on shell. If the neutrino 
is put on shell, then the one-loop integral is either over a kinematically forbidden momentum region or 
over a momentum region which is much smaller than $M$, according to the first remark above. 
If the light particles are put on shell, then, for we can neglect the 
momentum of the external Higgs boson, we have a situation equivalent to a vertex with three on-shell massless 
particles and the second remark above applies. The result is that diagrams 5) and 6) of figure~\ref{fig:new_Higgs_direct}
do not contribute to the CP asymmetry at the scale $M$, which is the result~\eqref{Higgs5}.

\subsection{Matching diagrams with gauge interactions}
\label{appgauge}
At order $F^4$ and at first order in the SM couplings, besides the Feynman diagrams with four-Higgs vertices 
computed in the previous section, also diagrams with a gauge boson can contribute.
We will compute them here.

\begin{figure}[ht]
\centering
\includegraphics[scale=0.5]{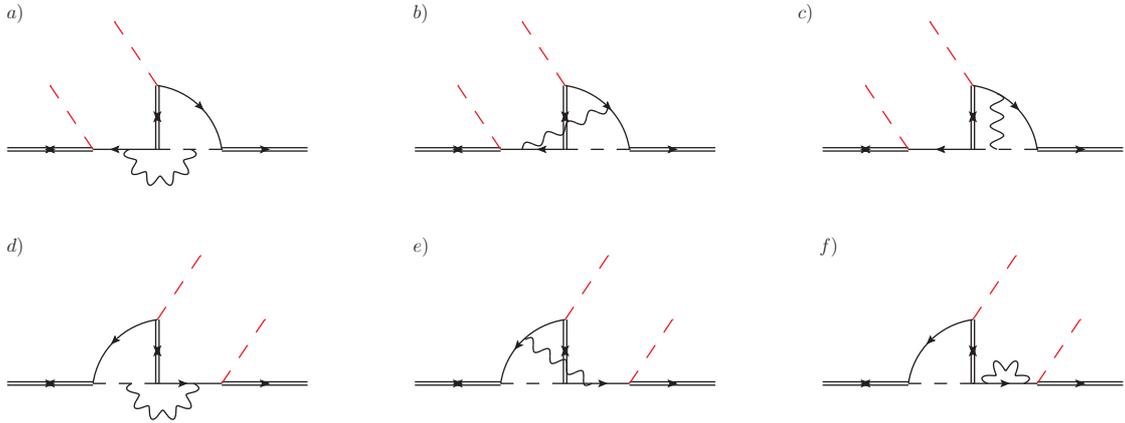}
\caption{If the incoming and outgoing Majorana neutrinos are conventionally chosen to be of type 1, 
then the displayed diagrams contribute to $a^\ell_{11}$ at order $F^4$ and at first order in the gauge couplings. 
The diagrams contribute also to $a^{\bar{\ell}}_{11}$ if cut through the anti-lepton.
Only diagrams proportional to $(F_{1}^{*} F_{2})^2$ are displayed.}
\label{fig:diagrams_1}
\end{figure}

By cutting this kind of diagrams we distinguish two different type of processes:
processes with a gauge boson in the final state or processes without a gauge boson in the final state. 
These being two distinct physical processes, we can compute them in different gauges. 
It is advantageous to adopt the Coulomb gauge in the first type of processes and the Landau gauge in the second one. The advantages are twofold.
First, with this choice of gauge we can neglect, for the purpose of matching the dimension-five operators~\eqref{aijoperator}
in the EFT, all diagrams with a gauge boson attached to an external Higgs boson leg.
The reason is that the coupling of the gauge boson with the Higgs boson is proportional to the momentum of the latter (see~\eqref{SMlag} and \eqref{SMCov}). 
If it depends on the external momentum, then the diagram will contribute to the matching of a higher-dimensional operator in the EFT, 
for the dimension-five operators do not contain derivatives. If it depends on the internal momentum then its contraction 
with the gauge boson propagator vanishes both in Landau gauge, if the gauge boson is uncut, and in Coulomb gauge, if the gauge boson is cut.
In the latter case, only transverse gauge bosons can be cut.
Second, the physical Coulomb gauge does not generate spurious singularities when the gauge boson is cut.

\begin{figure}[ht]
\centering
\includegraphics[scale=0.5]{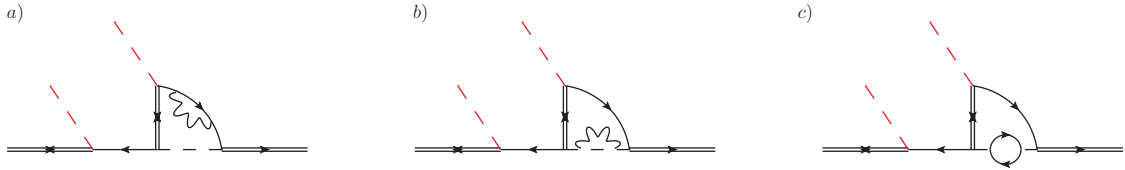}
\caption{Diagrams as in figure~\ref{fig:diagrams_1}. 
In diagram $c)$, the particles in the small loop coupled to a Higgs boson are a top-quark and a heavy-quark doublet.}
\label{fig:diagrams_2} 
\end{figure}

With the above choice of gauges, it is convenient to divide the remaining diagrams contributing to the matching of the dimension-five operators  
into the four sets shown in figures~\ref{fig:diagrams_1}, \ref{fig:diagrams_2}, \ref{fig:diagram_single} and~\ref{fig:fig_ind_new} 
for the leptonic contribution. After closer inspection, diagram~$c)$ in figure~\ref{fig:diagrams_1} turns out not to contribute to the CP asymmetry.
The diagram may be cut through the lepton propagator in two ways leaving in each case an uncut one-loop subdiagram.
The only cuts for these subdiagrams that are relevant for the matching (see discussion at the end of section~\ref{appHiggs}) 
give rise to two identical but opposite contributions (they differ only in the number of circled vertices), which cancel. 
We have checked the cancellation also by explicit calculation. 

\begin{figure}[ht]
\centering
\includegraphics[scale=0.42]{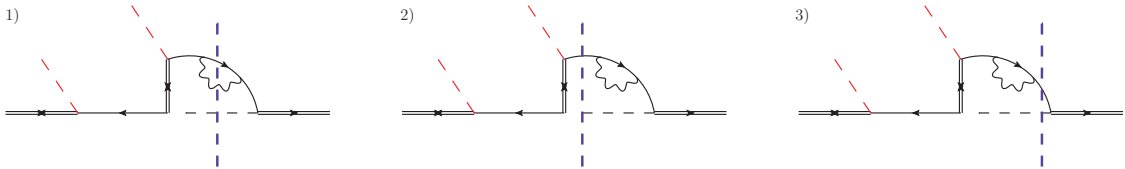}
\caption{Cuts on diagram $a)$ of figure~\ref{fig:diagrams_2}. 
The first cut does not contain any loop. The other two cut diagrams do contain a
remaining loop that however does not develop an imaginary part.}
\label{fig:cutexample} 
\end{figure}

We consider now the three diagrams in figure~\ref{fig:diagrams_2}. 
It turns out that these diagrams cannot introduce an additional complex phase, i.e., they do not develop 
an imaginary part of the loop amplitude, the quantity that we called ${\rm{Im}}(B)$ in section \ref{sec:zeroT}. 
In order to prove this statement, let us pick up diagram $a)$ in figure~\ref{fig:diagrams_2} 
and consider all possible cuts that put a lepton on shell. These are shown in figure~\ref{fig:cutexample}. 
The first cut does not contain any loop, hence it does not generate any additional complex phase besides the Yukawa couplings. 
In the second and third cut, in order to generate a complex phase, the remaining loop diagrams would need to develop an imaginary part. 
However, this is not the case since the (on-shell) incoming and outgoing particles in the loop and the particles in the loop itself are massless, 
a situation already discussed at the end of section~\ref{appHiggs}.
Therefore, also in this case, the diagram and its complex conjugate contribute with a term proportional to ${\rm{Re}}\left[(F_{1}^{*}F_{2})^2\right]$, 
which cancels eventually against the anti-leptonic width in the CP asymmetry. The same argument applies to both diagrams $b)$ and $c)$ in figure~\ref{fig:diagrams_2}
(as well as to diagrams with loops inserted in the external Higgs legs that we have not displayed).  
As an important consequence, there are not thermal corrections to the CP asymmetry of order~$T^2/M^2$ that are proportional  
to the top-Yukawa coupling, $\lambda_t$.

\begin{figure}[ht]
\centering
\includegraphics[scale=0.5]{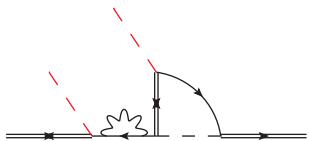}
\caption{Diagram as in figure~\ref{fig:diagrams_1}.}
\label{fig:diagram_single} 
\end{figure}

The diagram in figure~\ref{fig:diagram_single} does not contribute as well to the CP asymmetry. 
Indeed, once it has been cut in a way that the lepton and Higgs boson are on shell, 
what is left is a subdiagram with a vanishing imaginary part in Landau gauge. 
This has been shown by direct computation in~\cite{Biondini:2013xua}\footnote{
See figure~4, diagram 5), and eq.~(A.8) there.}.

\begin{figure}[ht]
\centering
\includegraphics[scale=0.5]{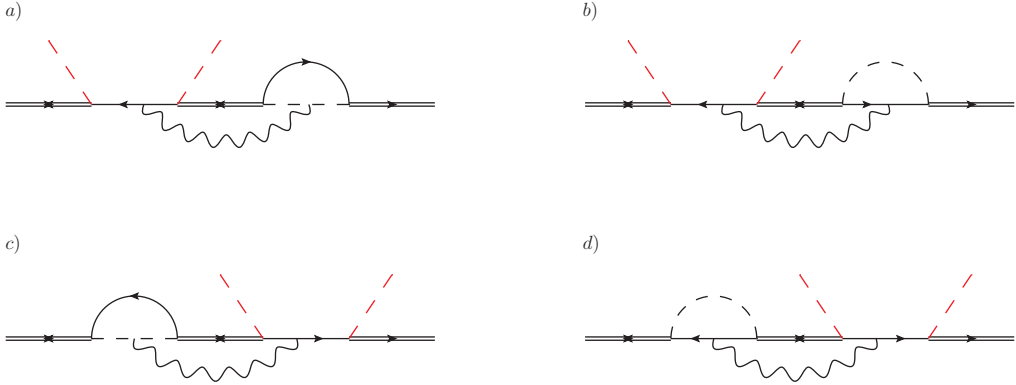}
\caption{Four diagrams that would be resonant without the gauge boson.
Only diagrams proportional to $(F_{1}^{*} F_{2})^2$ are displayed.}
\label{fig:fig_ind_new} 
\end{figure}

\begin{figure}[ht]
\centering
\includegraphics[scale=0.5]{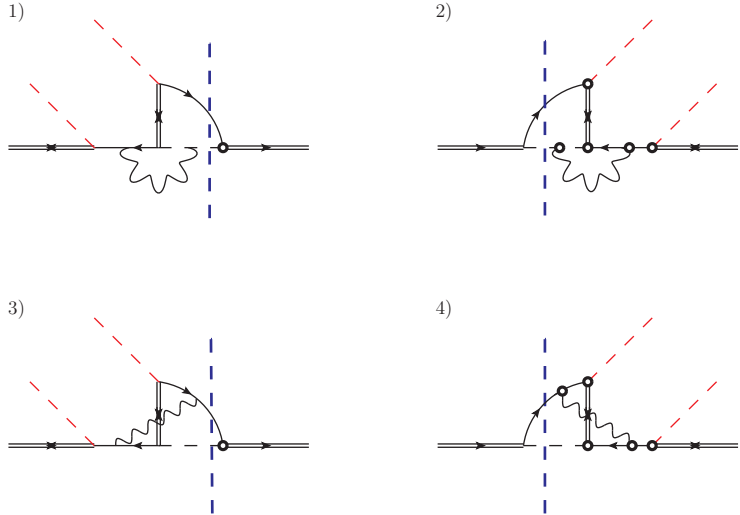}
\caption{On each raw we show the diagrams $a)$ and $b)$ of figure~\ref{fig:diagrams_1} 
together with their complex conjugates. Higgs bosons and leptons are cut.}
\label{appendix3}
\end{figure}

\begin{figure}[ht]
\centering
\includegraphics[scale=0.5]{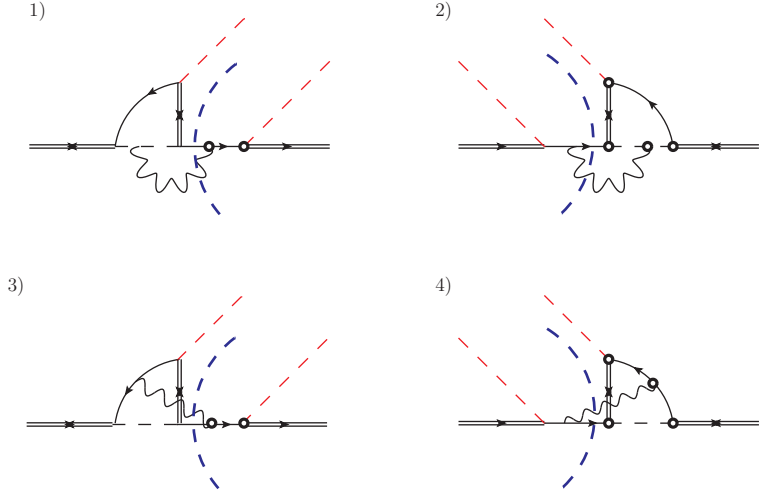}
\caption{On each raw we show the diagrams $d)$ and $e)$ of figure~\ref{fig:diagrams_1} 
together with their complex conjugates. Gauge bosons and leptons are cut.}
\label{appendix5} 
\end{figure}

\begin{figure}[ht]
\centering
\includegraphics[scale=0.5]{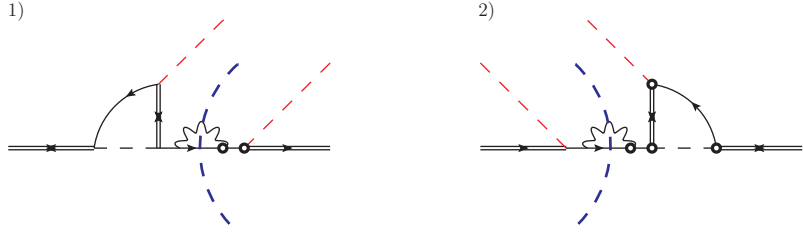}
\caption{Diagram $f)$ of figure~\ref{fig:diagrams_1} together with its complex conjugate. Gauge bosons and leptons are cut.}
\label{appendix6} 
\end{figure}

We compute now the part of $a^\ell_{11}$ relevant for the CP asymmetry coming from the diagrams 
of figure~\ref{fig:diagrams_1} that have not been already excluded on the basis of the previous arguments.
We organize the calculation as follows: first, we compute the cuts that go through the lepton 
but not the gauge boson, i.e., the gauge boson contributes only as a virtual particle in the loop, 
then we compute the cuts that go through both the lepton and the gauge boson.  
In figure~\ref{appendix3}, we show the cuts in the first case, 
whereas in figure~\ref{appendix5} and \ref{appendix6} we show them in the second one. 
On each raw we draw a diagram and its complex conjugate. 
As argued before, cuts that do not leave a loop uncut do not generate any additional complex phase 
and therefore do not contribute to the CP asymmetry. These cuts are not displayed.

We start with computing the cuts shown in figure~\ref{appendix3}. 
In Landau gauge, the result is 
\begin{eqnarray}
&&\!\!\!\!\!\!\!\! 
{\rm{Im}}\, (-i \mathcal{D}^{\ell}_{1,\hbox{\tiny fig.\ref{appendix3}}}) + {\rm{Im}}\, (-i \mathcal{D}^{\ell}_{2,\hbox{\tiny fig.\ref{appendix3}}}) = 0 \, , 
\label{boson1} \\
&&\!\!\!\!\!\!\!\! 
{\rm{Im}}\, (-i \mathcal{D}^{\ell}_{3,\hbox{\tiny fig.\ref{appendix3}}}) + {\rm{Im}}\, (-i \mathcal{D}^{\ell}_{4,\hbox{\tiny fig.\ref{appendix3}}}) = 
\nonumber\\
&&\hspace{2cm}
-  \frac{{\rm{Im}}\left[ (F^{*}_{1}F_{2})^2\right]}{(16 \pi)^2M}  \frac{3g^2+g'^{2}}{8} 
\left[ \ln 2 - \left( 1-\ln 2 \right) \frac{\Delta}{M}\right] \,\delta^{\mu \nu}  \delta_{mn} + \dots  \, ,  
\label{boson2}
\end{eqnarray}
where the superscript $\ell$ refers to having cut a lepton line. 
The dots stand for higher-order terms in the $\Delta/M$ expansion and for terms that do not contribute to the CP asymmetry.

We compute now cuts through gauge bosons. As argued at the beginning of this section, we can use for this kind of cuts a different gauge, 
namely the Coulomb gauge. The result for the cuts shown in figure~\ref{appendix5} reads
\begin{eqnarray}
&&\!\!\!\! {\rm{Im}}\, (-i \mathcal{D}^{\ell}_{1,\hbox{\tiny fig.\ref{appendix5}}}) + {\rm{Im}}\, (-i \mathcal{D}^{\ell}_{2,\hbox{\tiny fig.\ref{appendix5}}}) =
\nonumber \\
&& \hspace{1.2cm}
- \frac{{\rm{Im}}\left[ (F^{*}_{1}F_{2})^2\right]}{(16 \pi)^2M} \frac{3g^2+g'^2}{8}  \left(-1+\frac{\Delta}{M} \right) \delta^{\mu \nu} \delta_{mn}  + \dots  \, , 
\label{boson5} \\
&&\!\!\!\! {\rm{Im}}\, (-i \mathcal{D}^{\ell}_{3,\hbox{\tiny fig.\ref{appendix5}}}) + {\rm{Im}}\, (-i \mathcal{D}^{\ell}_{4,\hbox{\tiny fig.\ref{appendix5}}}) = 
\nonumber \\
&& \hspace{1.2cm}
- \frac{{\rm{Im}}\left[ (F^{*}_{1}F_{2})^2\right]}{(16 \pi)^2M} \frac{3g^2+g'^2}{4}  
\left[  \left(1-\ln 2 \right) +\left( 2-3 \ln 2\right) \frac{\Delta}{M} \right]   \,  \delta^{\mu \nu} \delta_{mn} + \dots \, .
\label{boson6}
\end{eqnarray}

Two more diagrams that contribute to the part of $a^\ell_{11}$ that matters for the CP asymmetry with the relevant cuts are shown in figure~\ref{appendix6}. 
They give
\begin{equation}
{\rm{Im}}\, (-i \mathcal{D}^{\ell}_{1,\hbox{\tiny fig.\ref{appendix6}}}) + {\rm{Im}}\, (-i \mathcal{D}^{\ell}_{2,\hbox{\tiny fig.\ref{appendix6}}}) =
- \frac{{\rm{Im}}\left[ (F^{*}_{1}F_{2})^2\right]}{(16 \pi)^2M} \frac{3g^2+g'^{2}}{8} \left(1-\frac{\Delta}{M} \right) \delta^{\mu \nu} \delta_{mn}  + \dots\,.
\label{boson7}
\end{equation}

Finally, we consider the diagrams shown in figure~\ref{fig:fig_ind_new}. Removing the gauge boson, these diagrams could become resonant and contribute 
to the indirect CP asymmetry discussed in section~\ref{sec:indirect}. Indeed their contribution is accounted for by the diagrams in 
the EFT shown in figure~\ref{fig:indirectEFT}. With the gauge bosons included these diagrams cannot become resonant when the gauge boson carries 
away an energy of order $M$ and, according to the definition adopted in this paper, they contribute to the direct CP asymmetry. 
Clearly they do contribute to the Wilson coefficients ${\rm{Im}} \,a_{II}^\ell$ and ${\rm{Im}} \,a_{II}^{\bar{\ell}}$.

\begin{figure}[ht]
\centering
\includegraphics[scale=0.5]{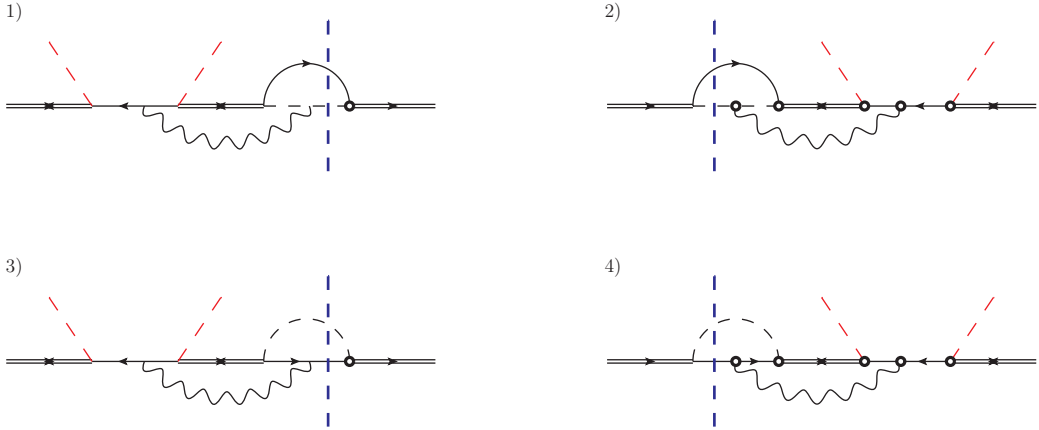}
\caption{On each raw we show the diagrams $a)$ and $b)$ of figure~\ref{fig:fig_ind_new} together with their complex conjugates.
Higgs bosons and leptons are cut.}
\label{appendix7}  
\end{figure}

As before, we start considering cuts through leptons and Higgs bosons. Only diagrams $a)$ and $b)$ of figure~\ref{fig:fig_ind_new}
may be cut in this way and contribute to the CP asymmetry. The diagrams and the relevant cuts are shown in figure~\ref{appendix7}. 
The result in Landau gauge reads
\begin{eqnarray}
&&\hspace{-1cm}
{\rm{Im}}\, (-i \mathcal{D}^{\ell}_{1,\hbox{\tiny fig.\ref{appendix7}}}) + {\rm{Im}}\, (-i \mathcal{D}^{\ell}_{2,\hbox{\tiny fig.\ref{appendix7}}}) =0     \, , 
\label{boson8} \\
&&\hspace{-1cm}
{\rm{Im}}\, (-i \mathcal{D}^{\ell}_{3,\hbox{\tiny fig.\ref{appendix7}}}) + {\rm{Im}}\, (-i \mathcal{D}^{\ell}_{4,\hbox{\tiny fig.\ref{appendix7}}}) = 
- \frac{{\rm{Im}}\left[ (F^{*}_{1}F_{2})^2\right]}{(16 \pi)^2M} \frac{g'^2}{4} \left(1-\frac{\Delta}{M} \right) \delta^{\mu \nu} \delta_{mn}  + \dots \, .
\label{boson9}
\end{eqnarray}

\begin{figure}[htb]
\centering
\includegraphics[scale=0.5]{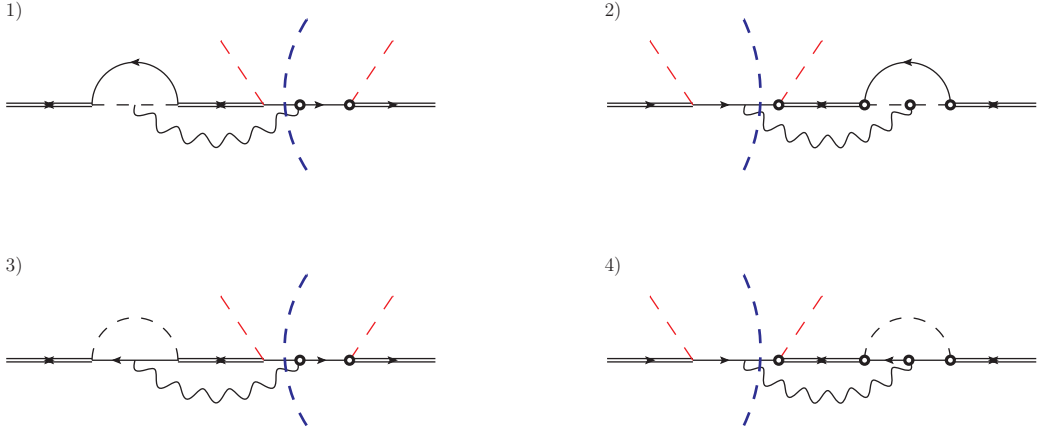}
\caption{On each raw we show the diagrams $c)$ and $d)$ of figure~\ref{fig:fig_ind_new} together with their complex conjugates.
Gauge bosons and leptons are cut.}
\label{appendix8} 
\end{figure}

On the other hand, only diagrams $c)$ and $d)$ of figure~\ref{fig:fig_ind_new}
may be cut through a lepton and a gauge boson. The diagrams and the relevant cuts are shown in figure~\ref{appendix8}. 
The result in Coulomb gauge reads
\begin{eqnarray}
&&\hspace{-1cm}
{\rm{Im}}\, (-i \mathcal{D}^{\ell}_{1,\hbox{\tiny fig.\ref{appendix8}}}) + {\rm{Im}}\, (-i \mathcal{D}^{\ell}_{2,\hbox{\tiny fig.\ref{appendix8}}}) =
- \frac{{\rm{Im}}\left[ (F^{*}_{1}F_{2})^2\right]}{(16 \pi)^2M} \frac{g'^2}{4}  \left(-1+\frac{\Delta}{M} \right) \delta^{\mu \nu} \delta_{mn}  + \dots  \, , 
\label{boson10} \\
&&\hspace{-1cm}
{\rm{Im}}\, (-i \mathcal{D}^{\ell}_{3,\hbox{\tiny fig.\ref{appendix8}}}) + {\rm{Im}}\, (-i \mathcal{D}^{\ell}_{4,\hbox{\tiny fig.\ref{appendix8}}}) = 
- \frac{{\rm{Im}}\left[ (F^{*}_{1}F_{2})^2\right]}{(16 \pi)^2M} \frac{g'^2}{4}  
\left(1-\frac{\Delta}{M} \right) \delta^{\mu \nu} \delta_{mn}  + \dots \, .
\label{boson11}
\end{eqnarray}
Note that the SU(2)$_L$ gauge bosons do not contribute to \eqref{boson9}-\eqref{boson11}.

Summing up all diagrams \eqref{Higgs1}-\eqref{boson11}, and comparing with the expression of 
the matrix element \eqref{B1} in the EFT, which is $({\rm{Im}} \,a_{11}^\ell/M) \delta^{\mu \nu} \delta_{mn}$ for the leptonic contribution 
and $({\rm{Im}} \,a_{11}^{\bar{\ell}}/M) \delta^{\mu \nu} \delta_{mn} $ for the anti-leptonic one, we obtain~\eqref{match1}.
The expression for the Wilson coefficient involving the Majorana neutrino of type 2 can be inferred from the above 
results after the substitutions $F_1 \leftrightarrow F_2$, $M \to M_2$ and $\Delta \to -\Delta$ in~\eqref{Higgs1}-\eqref{boson11} or just in~\eqref{match1}. 
The result, in terms of the lightest neutrino mass, $M$, has been written in~\eqref{match2}.
That the above substitutions work follows from the fact that the real transition from a heavier neutrino of type 2 to a lighter neutrino of type 1, 
which is a decay channel absent in the case of neutrinos of type 1, is a process accounted for by the EFT (see section~\ref{sec:direct2}), 
and, therefore, it does not contribute to the matching. 
In fact, the energy emitted in such a transition is of order~$\Delta$; 
this is, in the nearly degenerate case considered in this work, much smaller than~$M$.

\subsection{Matching the flavoured asymmetry}
\label{AppendixC}
There are diagrams contributing to the matching coefficients ${\rm{Im}} \,a_{II}^\ell$ and ${\rm{Im}} \,a_{II}^{\bar{\ell}}$ 
that are relevant only for the flavoured CP asymmetry.
These are diagrams involving only lepton (or anti-lepton) propagators. 
They could contribute to the CP asymmetry with terms proportional to ${\rm{Im}}\left[(F_1F_2^{*})(F^*_{f1}F_{f2})\right]$.
Clearly such terms vanish in the unflavoured case. 
Here we examine these diagrams and find that they do not contribute.

\begin{figure}[ht]
\centering
\includegraphics[scale=0.53]{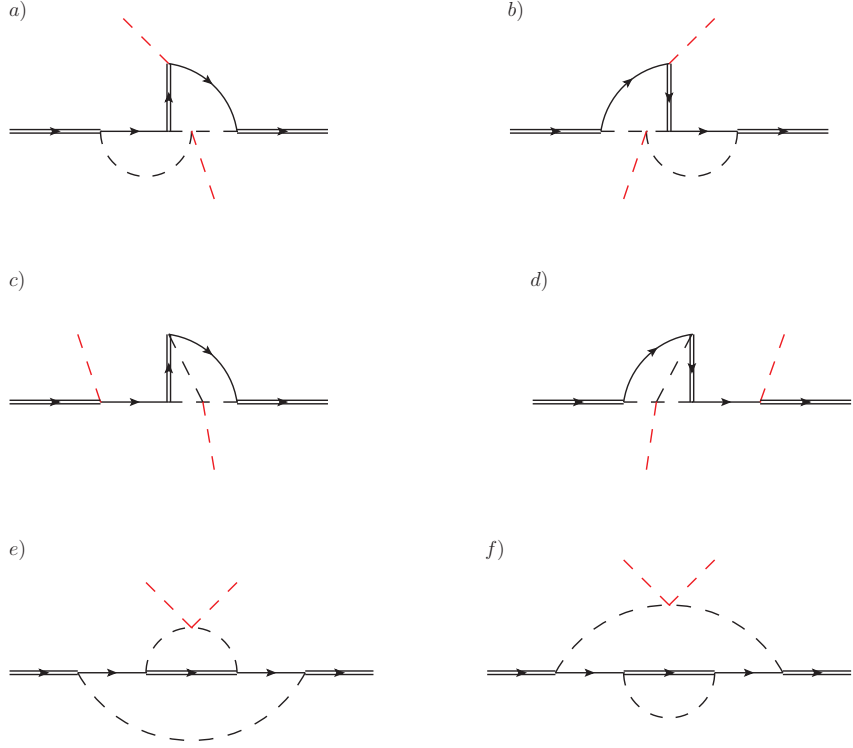}
\caption{Diagrams contributing to the matching coefficients \eqref{match1} and \eqref{match2} involving the four-Higgs coupling.
 Diagrams $a)$-$d)$ may be inferred from the diagrams of figure~\ref{fig:new_Higgs_direct} by changing an anti-lepton line in a lepton line.
 The topologies of diagrams $e)$ and $f)$ are relevant only for the flavoured case. 
 We display only diagrams that admit leptonic cuts.}
\label{fig:flavor_Higgs} 
\end{figure} 

\begin{figure}[ht]
\centering
\includegraphics[scale=0.53]{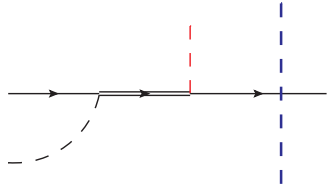}
\caption{The blue dashed line on the right is the cut, the red central dashed line is an external Higgs boson whose momentum can be set to zero and
the black dashed line on the left may identify a Higgs boson in a loop or an external one.}
\label{fig:flavour_cut} 
\end{figure} 

\begin{figure}[ht]
\centering
\includegraphics[scale=0.53]{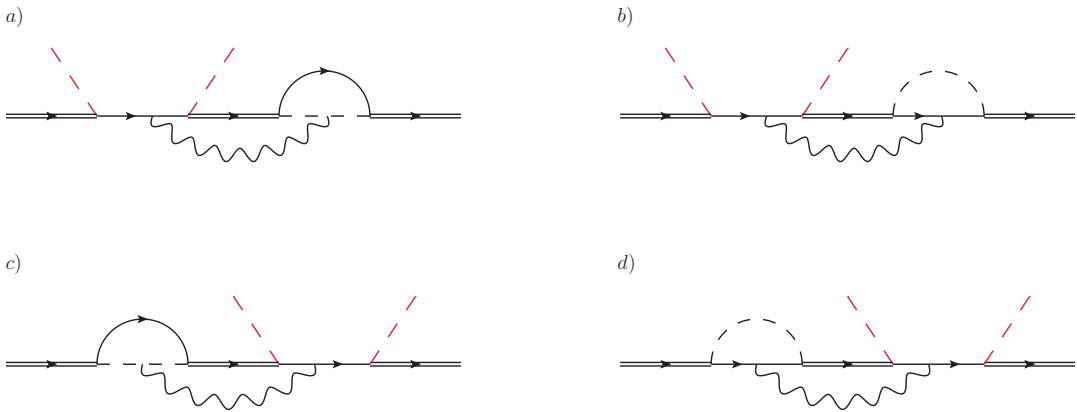}
\caption{Diagrams contributing to the matching coefficients \eqref{match1} and \eqref{match2} involving gauge couplings.
The diagrams may be inferred from the diagrams of figure~\ref{fig:fig_ind_new} by changing an anti-lepton line in a lepton line.
We display only diagrams that admit leptonic cuts.}
\label{fig:flavor_gauge} 
\end{figure}

\begin{figure}[ht]
\centering
\includegraphics[scale=0.53]{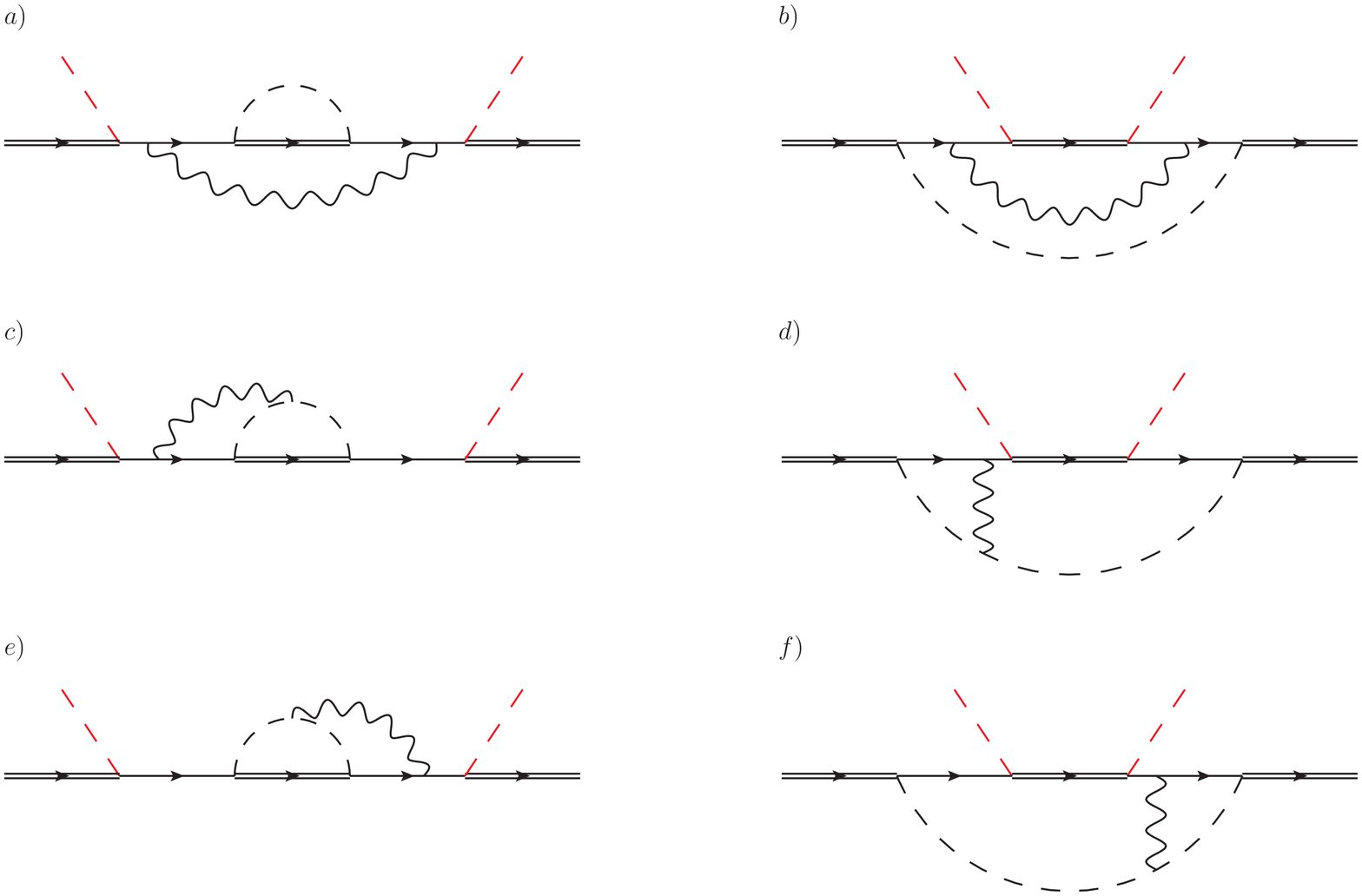}
\caption{Diagrams contributing to the matching coefficients \eqref{match1} and \eqref{match2} involving gauge couplings.
The topologies of these diagrams are relevant only for the flavoured case. 
We display only diagrams that admit leptonic cuts.}
\label{fig:flavor_gauge_rainbow} 
\end{figure}

We may divide these diagrams into two classes: diagrams that involve the four-Higgs coupling, shown in figure~\ref{fig:flavor_Higgs}, 
and diagrams involving gauge couplings, shown in figures~\ref{fig:flavor_gauge} and~\ref{fig:flavor_gauge_rainbow}.
Let us consider diagram $a)$ of figure~\ref{fig:flavor_Higgs}. If we cut the lepton in the loop on the right, then the cut gives rise to 
the Feynman subdiagram shown in figure~\ref{fig:flavour_cut}. This is proportional to ($\ell^\mu$ is the momentum of the lepton) 
\begin{equation}
\delta(\ell^2)\slashed{\ell} \, P_R \, \frac{\slashed{\ell}+M_J}{\ell^2-M_J^2+i\epsilon}\, P_L = P_L \, \delta(\ell^2) \ell^2 \frac{1}{\ell^2-M_J^2+i\epsilon} = 0,
\label{flavour_identity}
\end{equation} 
and therefore vanishes.\footnote{
The corresponding Feynman subdiagram of 1) in figure~\ref{fig:new_Higgs_direct} involves a neutrino propagator of the type \eqref{A2}
and an anti-lepton on the left. Hence it is proportional to 
$$
\delta(\ell^2)\slashed{\ell} \, P_R \, \frac{\slashed{\ell}+M_J}{\ell^2-M_J^2+i\epsilon}\, P_R = 
P_L \, \delta(\ell^2) \slashed{\ell} M_J \frac{1}{\ell^2-M_J^2+i\epsilon} \neq 0.
$$
} 
If we cut the lepton in the loop on the left, then we need the imaginary part of the remaining (uncut) loop on the right.
The imaginary part of the loop on the right may be computed by considering all its possible cuts. Those include cuts through the lepton, which 
vanish according to the above argument, cuts through the Higgs-boson propagator, which vanish because they involve three massless on-shell particles 
entering the same vertex, and cuts through the Majorana-neutrino propagator, which are either kinematically forbidden or involve momenta 
of order $\Delta$ that are accounted for by the EFT (for more details see the discussion at the end of section~\ref{appHiggs}).

The same arguments may be applied to all remaining diagrams shown in figures~\ref{fig:flavor_Higgs}, \ref{fig:flavor_gauge} and \ref{fig:flavor_gauge_rainbow}.
In particular, for many of them the argument based on the identity~\eqref{flavour_identity} is crucial.
The identity~\eqref{flavour_identity} is relevant only for the flavoured case.

\section{The $T/M$ expansion}
\label{AppendixD}
In the paper, we have computed the thermal corrections to the neutrino CP asymmetry as an expansion in the SM couplings and in $T/M$.
The production rate for heavy Majorana neutrinos has been computed in a similar fashion in~\cite{Salvio:2011sf,Laine:2011pq,Biondini:2013xua}. 
Up to the order to which it is known, the expansion in $T/M$ is well behaved, i.e., for reasonably small values of $T/M$ it converges. 

\begin{figure}[ht]
\centering
\includegraphics[scale=0.8]{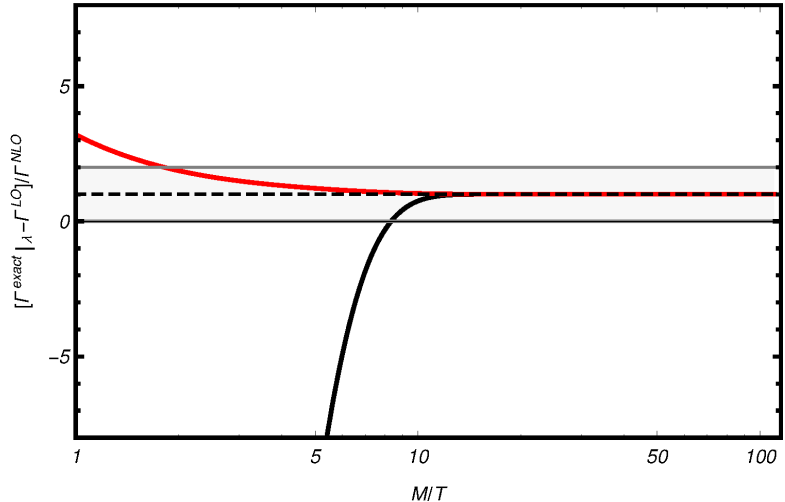}
\caption{The black line shows the difference between the exact neutrino production rate up to order $\lambda$ (top-Yukawa and gauge couplings set to zero) 
and the neutrino production rate at leading order in $T/M$ divided by the neutrino production rate at next-to-leading order in $T/M$.
The red line is as above but with the leading-order neutrino production rate multiplied by $(1 + n_B(M/2) - n_F(M/2))$. The neutrino is taken at rest.
The one-loop running four-Higgs coupling, $\lambda$, is taken $\lambda(10^7~{\rm GeV}) \approx 0.02$ ($\lambda(125~{\rm GeV}) \approx 0.126$)~\cite{Rose15}.
}
\label{Gammaexpansion} 
\end{figure}

Despite the above fact, it has been remarked in~\cite{Laine:2013lka} that, 
when comparing the production rate for heavy Majorana neutrinos in the $T/M$ expansion with the exact result, 
which is known at leading order in the SM couplings, the two results overlap only at very small values of $T/M$, i.e., values around $1/10$ or smaller.
In the same work, it has been also noticed that for values of $T/M$ larger than $1/10$ 
not only the discrepancy between the exact and the approximate result appears larger than the last known term in the expansion, but also of opposite sign.
The situation is well illustrated by the black curve in figure~\ref{Gammaexpansion}. 
It shows the difference between the exact neutrino production rate at order $\lambda$ (top-Yukawa and gauge couplings are set to zero) taken from~\cite{Laine:2013lka}
and the neutrino production rate at leading order in $T/M$ divided by the neutrino production rate at next-to-leading order in $T/M$. 
At next-to-leading order in $T/M$ the production rate depends only on the SM coupling $\lambda$.

The same behaviour may potentially show up also for the CP asymmetry, although in this case the exact result is unknown. 
For this reason, in the rest of the appendix we will clarify the origin of this behaviour 
and devise a strategy to improve the expansion in $T/M$ in such a way that it overlaps with the exact result for reasonably small, 
not only very small, values of $T/M$. We will say that the expansion overlaps with the exact result if the discrepancy between the exact 
and the approximate result is not larger than the last known term in the expansion.

The problem is rather general. In the form we have it here, it happens when dealing with a double expansion 
where one of the expansion parameters is much smaller than the other one.
In our case $\lambda$ is much smaller than $T/M$ for a relatively wide range of temperatures.
Under this circumstance, exponentially suppressed terms of the type $e^{-M/T}$ 
may become numerically as large as next-to-leading order terms of the type $\lambda\,(T/M)^2$.
In fact $e^{-M/T}$ is larger than or very close to $\lambda\,(T/M)^2$ for $T/M \simg 1/8$.
One should recall that exponentially suppressed terms vanish in any analytic expansion.

The solution of the problem consists in keeping exponentially suppressed terms in the not-so-small parameter 
at leading order in the small-parameter expansion. 
In our case, this amounts at keeping terms of the type $e^{-M/T}$ in the computation of the neutrino observables at zeroth-order in the SM couplings. 
Let us illustrate how this works in the case of the neutrino production rate.
The relevant diagram is the self-energy diagram 1) of figure~\ref{Fig2}, which, in the following, we will call $\Pi$.
The neutrino production rate is proportional to the retarded self energy, $\Pi_R$.
In turn, the retarded self energy may be written as $\Pi_R = \Pi_{11} + \Pi_{12}$, 
where $\Pi_{11}$ is the self energy when the initial and final neutrinos are on the physical branch of the Keldysh contour, 
and $\Pi_{12}$ is the self energy when the initial neutrino is on the physical branch whereas the final neutrino 
is on the complex branch of the Keldysh contour~\cite{Bellac,Thoma:2000dc}. 
The ``12'' component of a heavy-particle propagator vanishes exponentially in the heavy-mass limit~\cite{Brambilla:2008cx}. 
For this reason we did not need to consider $\Pi_{12}$ in~\cite{Biondini:2013xua}. 
But we need to consider it here if we want to keep exponentially suppressed terms. 
Cutting $\Pi_{11}$ and keeping the thermal distributions of the lepton and Higgs boson gives for a neutrino at rest 
$\Pi_{11} = \left[T=0~{\rm result}\right] \times (1+n_B(M/2)) (1-n_F(M/2))$, where $n_B(E)=1/(e^{E/T}-1)$ and $n_F(E)=1/(e^{E/T}+1)$ are the Bose and Fermi distributions respectively.
Cutting $\Pi_{12}$ gives $\Pi_{12} = \left[ T=0~{\rm result} \right]  \times n_B(M/2) n_F(M/2)$.
Summing the two contributions gives $\Pi_R = \left[ T=0~{\rm result} \right]  \times (1 + n_B(M/2) - n_F(M/2))$.
Hence, we can improve the neutrino production rate at leading order in the SM coupling by multiplying the $T=0$ result by 
\begin{equation}
1 + n_B(M/2) - n_F(M/2) \approx 1 + 2\,e^{-M/T}  + ...\;,
\label{expimprovement}
\end{equation}
which amounts at keeping (at least) terms of the type $e^{-M/T}$ in the expansion of the Bose and Fermi distributions for $M \gg T$.

In figure~\ref{Gammaexpansion} the red curve shows the difference between the exact neutrino production rate at order $\lambda$ 
(top-Yukawa and gauge couplings set to zero) and the neutrino production rate at leading order in $T/M$ 
multiplied by $(1 + n_B(M/2) - n_F(M/2))$ divided by the neutrino production rate at next-to-leading order in $T/M$.
The grey band shows the region where the discrepancy between the exact production rate and the next-to-leading order one 
is not larger than the next-to-leading order one. 
We see that now the curve is in the grey band for $T/M \siml 1/2$.
Moreover, higher-order corrections in $T/M$ do not change the sign of the next-to-leading order correction.
The result is consistent with our understanding of the problem and in fact provides a simple way to solve it.

This computational scheme could be also implemented in the case of the CP asymmetry. 
For the direct CP asymmetry, the leading-order diagrams are in this case given by the two-loop diagrams shown in figure~\ref{Fig2}.
Because we are cutting them and taking the imaginary parts of the remaining one-loop subdiagrams, 
exponentially suppressed contributions can be computed straightforwardly taking into account the combinatorics of all possible physical 
and unphysical degrees of freedom contributing to $\Pi_{11}$ and $\Pi_{12}$ at two loops.
A computation along this line is in~\cite{Garny:2010nj}.
For the indirect CP asymmetry, the computation may be done in the EFT, whose parameters are the thermal decay widths and masses.
The exponential improvement of the widths has been discussed in the previous paragraphs. 

Finally, we comment about the neutrino momentum $k$. Strictly speaking the non-relativistic expansion is an expansion in $T/M$ and $k/M$ 
and is as good as these two parameters are small. If $k$ is chosen to be equal to $T$ or smaller, 
as we did in figure~\ref{Gammaexpansion}, then $T/M$ is the relevant expansion parameter.
But if $k = 2T$, $k= 3T$, ... then this is $k/M$.
In particular, one has to expect (naively) the exact result to overlap with the result of the perturbative series 
at temperature $2$, $3$, ... times smaller than one would have for $k \le T$.


\newpage


\begin{thebibliography}{99}
\bibitem{Larson:2010gs}
  D.~Larson, J.~Dunkley, G.~Hinshaw, E.~Komatsu, M.~R.~Nolta, C.~L.~Bennett, B.~Gold and M.~Halpern {\it et al.},
  Astrophys.\ J.\ Suppl.\  {\bf 192} (2011) 16
  [arXiv:1001.4635 [astro-ph.CO]].
  
\bibitem{Iocco:2008va}
  F.~Iocco, G.~Mangano, G.~Miele, O.~Pisanti and P.~D.~Serpico,
  Phys.\ Rept.\  {\bf 472} (2009) 1
  [arXiv:0809.0631 [astro-ph]].
  
\bibitem{Fukugita:1986hr}
  M.~Fukugita and T.~Yanagida,
  Phys.\ Lett.\ B {\bf 174} (1986) 45.

\bibitem{Kuzmin:1985mm}
  V.~A.~Kuzmin, V.~A.~Rubakov and M.~E.~Shaposhnikov,
  Phys.\ Lett.\ B {\bf 155} (1985) 36.

\bibitem{Laine:2013lka}
  M.~Laine,
  JHEP {\bf 1308} (2013) 138
  [arXiv:1307.4909 [hep-ph]].
 
\bibitem{Salvio:2011sf}
  A.~Salvio, P.~Lodone and A.~Strumia,
  JHEP {\bf 1108} (2011) 116
  [arXiv:1106.2814 [hep-ph]].

\bibitem{Laine:2011pq}
  M.~Laine and Y.~Schr\"oder,
  JHEP {\bf 1202} (2012) 068
  [arXiv:1112.1205 [hep-ph]].

\bibitem{Biondini:2013xua}
  S.~Biondini, N.~Brambilla, M.~A.~Escobedo and A.~Vairo,
  JHEP {\bf 1312} (2013) 028
  [arXiv:1307.7680].

\bibitem{Sakharov:1967dj}
  A.~D.~Sakharov,
  Pisma Zh.\ Eksp.\ Teor.\ Fiz.\  {\bf 5} (1967) 32
   [JETP Lett.\  {\bf 5} (1967) 24]
   [Sov.\ Phys.\ Usp.\  {\bf 34} (1991) 392]
   [Usp.\ Fiz.\ Nauk {\bf 161} (1991) 61].

\bibitem{Kolb:1979ui}
  E.~W.~Kolb and S.~Wolfram,
  Phys.\ Lett.\ B {\bf 91} (1980) 217;
  Nucl.\ Phys.\ B {\bf 172} (1980) 224
   [Nucl.\ Phys.\ B {\bf 195} (1982) 542].

\bibitem{Liu:1993tg}
  J.~Liu and G.~Segre,
  Phys.\ Rev.\ D {\bf 48} (1993) 4609
  [hep-ph/9304241].

\bibitem{Covi:1996wh}
  L.~Covi, E.~Roulet and F.~Vissani,
  Phys.\ Lett.\ B {\bf 384} (1996) 169
  [hep-ph/9605319].

\bibitem{Flanz:1996fb}
  M.~Flanz, E.~A.~Paschos, U.~Sarkar and J.~Weiss,
  Phys.\ Lett.\ B {\bf 389} (1996) 693
  [hep-ph/9607310].

\bibitem{Buchmuller:1997yu}
  W.~Buchm\"uller and M.~Pl\"umacher,
  Phys.\ Lett.\ B {\bf 431} (1998) 354
  [hep-ph/9710460].

\bibitem{Garny:2011hg}
  M.~Garny, A.~Kartavtsev and A.~Hohenegger,
  Annals Phys.\  {\bf 328} (2013) 26
  [arXiv:1112.6428 [hep-ph]].

\bibitem{Garbrecht:2011aw}
  B.~Garbrecht and M.~Herranen,
  Nucl.\ Phys.\ B {\bf 861} (2012) 17
  [arXiv:1112.5954 [hep-ph]].

\bibitem{Pilaftsis:2003gt}
  A.~Pilaftsis and T.~E.~J.~Underwood,
  Nucl.\ Phys.\ B {\bf 692} (2004) 303
  [hep-ph/0309342].

\bibitem{Pilaftsis:1998pd}
  A.~Pilaftsis,
  Int.\ J.\ Mod.\ Phys.\ A {\bf 14} (1999) 1811
  [hep-ph/9812256].  
 
\bibitem{Covi:1997dr}
  L.~Covi, N.~Rius, E.~Roulet and F.~Vissani,
  Phys.\ Rev.\ D {\bf 57} (1998) 93
  [hep-ph/9704366].

\bibitem{Giudice:2003jh}
  G.~F.~Giudice, A.~Notari, M.~Raidal, A.~Riotto and A.~Strumia,
  Nucl.\ Phys.\ B {\bf 685} (2004) 89
  [hep-ph/0310123].

\bibitem{Garny:2010nj}
  M.~Garny, A.~Hohenegger and A.~Kartavtsev,
  Phys.\ Rev.\ D {\bf 81} (2010) 085028
  [arXiv:1002.0331 [hep-ph]].

\bibitem{Anisimov:2010dk}
  A.~Anisimov, W.~Buchm\"uller, M.~Drewes and S.~Mendizabal,
  Annals Phys.\  {\bf 326} (2011) 1998
   [Annals Phys.\  {\bf 338} (2011) 376]
  [arXiv:1012.5821 [hep-ph]].

\bibitem{Kiessig:2011fw}
  C.~Kiessig and M.~Pl\"umacher,
  JCAP {\bf 1207} (2012) 014
  [arXiv:1111.1231 [hep-ph]].

\bibitem{Biondinihier}
  S.~Biondini, N.~Brambilla and A.~Vairo, TUM-EFT 73/15, in preparation.

\bibitem{Asaka:2006rw}
  T.~Asaka, M.~Laine and M.~Shaposhnikov,
  JHEP {\bf 0606} (2006) 053
  [hep-ph/0605209].

\bibitem{Kniehl:1996bd}
  B.~A.~Kniehl and A.~Pilaftsis,
  Nucl.\ Phys.\ B {\bf 474} (1996) 286
  [hep-ph/9601390].

\bibitem{Anisimov:2005hr}
  A.~Anisimov, A.~Broncano and M.~Pl\"umacher,
  Nucl.\ Phys.\ B {\bf 737} (2006) 176
  [hep-ph/0511248].

\bibitem{Fong:2013wr}
  C.~S.~Fong, E.~Nardi and A.~Riotto,
  Adv.\ High Energy Phys.\  {\bf 2012} (2012) 158303
  [arXiv:1301.3062 [hep-ph]].

\bibitem{Denner:2014zga}
  A.~Denner and J.~N.~Lang,
  Eur.\ Phys.\ J.\ C {\bf 75} (2015) 8,  377
  [arXiv:1406.6280 [hep-ph]].

\bibitem{Pilaftsis:1997dr}
  A.~Pilaftsis,
  Nucl.\ Phys.\ B {\bf 504} (1997) 61
  [hep-ph/9702393].

\bibitem{Pilaftsis:1997jf}
  A.~Pilaftsis,
  Phys.\ Rev.\ D {\bf 56} (1997) 5431
  [hep-ph/9707235].

\bibitem{Dev:2014laa}
  P.~S.~Bhupal Dev, P.~Millington, A.~Pilaftsis and D.~Teresi,
  Nucl.\ Phys.\ B {\bf 886} (2014) 569
  [arXiv:1404.1003 [hep-ph]].

\bibitem{Frossard:2012pc}
  T.~Frossard, M.~Garny, A.~Hohenegger, A.~Kartavtsev and D.~Mitrouskas,
  Phys.\ Rev.\ D {\bf 87} (2013) 8,  085009
  [arXiv:1211.2140 [hep-ph]].

\bibitem{Garbrecht:2014aga}
  B.~Garbrecht, F.~Gautier and J.~Klaric,
  JCAP {\bf 1409} (2014) 09,  033
  [arXiv:1406.4190 [hep-ph]].

\bibitem{Weldon:1982bn}
  H.~A.~Weldon,
  Phys.\ Rev.\ D {\bf 26} (1982) 2789.

\bibitem{Nardi:2005hs}
  E.~Nardi, Y.~Nir, J.~Racker and E.~Roulet,
  JHEP {\bf 0601} (2006) 068
  [hep-ph/0512052].  
  
\bibitem{Nardi:2006fx}
  E.~Nardi, Y.~Nir, E.~Roulet and J.~Racker,
  JHEP {\bf 0601} (2006) 164
  [hep-ph/0601084].  
 
\bibitem{Campbell:1992jd}
  B.~A.~Campbell, S.~Davidson, J.~R.~Ellis and K.~A.~Olive,
  Phys.\ Lett.\ B {\bf 297} (1992) 118
  [hep-ph/9302221].  

\bibitem{Cline:1993bd}
  J.~M.~Cline, K.~Kainulainen and K.~A.~Olive,
  Phys.\ Rev.\ D {\bf 49} (1994) 6394
  [hep-ph/9401208].  
   
\bibitem{Davidson:2008bu}
  S.~Davidson, E.~Nardi and Y.~Nir,
  Phys.\ Rept.\  {\bf 466} (2008) 105
  [arXiv:0802.2962 [hep-ph]].  
 
\bibitem{Blanchet:2006be}
  S.~Blanchet and P.~Di Bari,
  JCAP {\bf 0703} (2007) 018
  [hep-ph/0607330].  

\bibitem{DeSimone:2006nrs}
  A.~De Simone and A.~Riotto,
  JCAP {\bf 0702} (2007) 005
  [hep-ph/0611357].  

\bibitem{Bodeker:2013qaa}
  D.~B\"odeker and M.~W\"ormann,
  JCAP {\bf 1402} (2014) 016
  [arXiv:1311.2593 [hep-ph]].

\bibitem{Rose15}
  L.~delle Rose, private communication.

\bibitem{Buttazzo:2013uya}
  D.~Buttazzo, G.~Degrassi, P.~P.~Giardino, G.~F.~Giudice, F.~Sala, A.~Salvio and A.~Strumia,
  JHEP {\bf 1312} (2013) 089
  [arXiv:1307.3536 [hep-ph]].

\bibitem{Cutkosky:1960sp}
  R.~E.~Cutkosky,
  J.\ Math.\ Phys.\  {\bf 1} (1960) 429.

\bibitem{Remiddi:1981hn}
  E.~Remiddi,
  Helv.\ Phys.\ Acta {\bf 54} (1982) 364.
  
\bibitem{Bellac}
  M.~Le Bellac,
  Quantum and Statistical Field Theory, (Oxford University Press, 1991).
  
\bibitem{Thoma:2000dc}
  M.~H.~Thoma,
  hep-ph/0010164.
  
\bibitem{Brambilla:2008cx}
  N.~Brambilla, J.~Ghiglieri, A.~Vairo and P.~Petreczky,
  Phys.\ Rev.\ D {\bf 78} (2008) 014017
  [arXiv:0804.0993 [hep-ph]].

\end{thebibliography}
\end{document}